\documentclass[aps,prc,twocolumn,showpacs,preprintnumbers,
               nofootinbib,float,superscriptaddress,longbibliography]{revtex4-1}
\usepackage{graphicx, fancybox}
\usepackage{amsmath,amssymb}
\usepackage[colorlinks=true, pdfstartview=FitV, linkcolor=red, citecolor=blue, urlcolor=blue]{hyperref}
\usepackage{color}
\usepackage{xcolor}
\usepackage{soul}
\usepackage{array}
\usepackage{url}
\usepackage[utf8]{inputenc}
\usepackage{multirow}
\usepackage{graphicx}

\graphicspath{{figs/}}

\usepackage[normalem]{ulem}

\begin{document}

\title{Exploring theoretical uncertainties in the hydrodynamic description of relativistic heavy-ion collisions}

\author{Cheng Chiu}
\email{cchiu22@cranbrook.edu}
\affiliation{Cranbrook Kingswood Upper School, Bloomfield Hills, Michigan, 48304, USA}

\author{Chun Shen}
\email{chunshen@wayne.edu}
\affiliation{Department of Physics and Astronomy, Wayne State University, Detroit, Michigan, 48201, USA}
\affiliation{RIKEN BNL Research Center, Brookhaven National Laboratory, Upton, NY 11973, USA}

\begin{abstract}
We explore theoretical uncertainties in the hydrodynamic description of relativistic heavy-ion collisions by examining the full non-linear causality conditions and quantifying the second-order transport coefficients' role on flow observables. The causality conditions impose physical constraints on the maximum allowed values of inverse Reynolds numbers during the hydrodynamic evolution. Including additional second-order gradient terms in the Denicol-Niemi-Moln\'{a}r-Rischke (DNMR) theory significantly shrinks the casual regions compared to those in the Israel-Stewart hydrodynamics. For Au+Au collisions, we find the variations of flow observables are small with and without imposing the necessary causality conditions, suggesting a robust extraction of the Quark-Gluon Plasma's transport coefficients in previous model-to-data comparisons. However, sizable sensitivity is present in small p+Au collisions, which poses challenges to study the small systems' collectivity.
\end{abstract}

{\maketitle}

\section{Introduction}

Relativistic viscous hydrodynamics has been the most successful model to provide a quantitative macroscopic description of heavy-ion collisions' dynamics at high energies \cite{Romatschke:2009im,Heinz:2013th,Gale:2013da,Yan:2017ivm,Florkowski:2017olj,Romatschke:2017ejr,Shen:2020mgh}. It is an efficient and effective phenomenological framework to extract many-body properties of nuclear matter under extremely hot and dense conditions.
As the QCD macroscopic properties emerge from the interactions among quarks and gluons, the QGP transport coefficients can be extracted from the comparisons between the hydrodynamic modeling and experimental data \cite{Romatschke:2007mq, Song:2010mg, Shen:2011kn, Gale:2012rq, Ryu:2015vwa, Shen:2015msa}.
Hydrodynamic simulations also provide detailed space-time evolution of relativistic heavy-ion collisions for studying the modification of rare probes under a hot nuclear environment, such as enhanced thermal electromagnetic radiation and suppression of QCD jets \cite{Majumder:2011uk, Burke:2013yra, Shen:2015nto, Shen:2016odt, Paquet:2015lta, Vujanovic:2019yih, Kumar:2019uvu, Tachibana:2020mtb}.

Most phenomenological simulations ensure causality on the linear level by choosing the relaxation times for the shear and bulk viscosity to be larger than the linear causality conditions \cite{Hiscock:1983zz, Olson:1989ey, Pu:2009fj, Huang:2010sa}. These conditions are ``static'', i.e., the causality bounds purely depend on the transport coefficients as functions of temperature and chemical potentials. There were attempts to go beyond the linear regime in restricted in 1+1 dimensions or with strong symmetry conditions \cite{Denicol:2008ha, Floerchinger:2017cii}. Recently, the full non-linear causality conditions were derived in Ref.~\cite{Bemfica:2020xym} for the Israel-Stewart (IS) and Denicol-Niemi-Moln\'{a}r-Rischke (DNMR) theories \cite{Israel:1976tn, Israel:1979wp, Muller:1967zza, Denicol:2012cn}. These more robust causality conditions directly involve the dynamically evolved shear stress tensor and bulk viscous pressure. Therefore, the equation of motion of IS and DNMR hydrodynamics alone can not guarantee that system will always stay within the causality region during the evolution. Causalilty conditions need to be examined locally in each fluid cell throughout the entire evolution.

The causality conditions can serve as physics constraints on the sizes of the shear stress tensor $\pi^{\mu\nu}$ and bulk viscous pressure $\Pi$. The constraints on their sizes were introduced as numerical regulators to stabilize the event-by-event numerical simulations \cite{Shen:2014vra, Denicol:2018wdp, Schenke:2020mbo}. The causality conditions provide us with useful theoretical guidance on these regulators' numerical choices in relativistic hydrodynamic simulations.

Recently, Bayesian Inference techniques were applied to systematically constrain multiple model parameters with various experimental measurements \cite{Bernhard:2016tnd, Moreland:2018gsh, Bernhard:2019bmu, Everett:2020yty, Everett:2020xug, Nijs:2020roc, Nijs:2020ors}. The causality conditions can provide additional constraints on the prior ranges for the first-order and second-order transport coefficients. At the same time, ensuring the full non-linear causality conditions will reduce the theoretical uncertainty from the choices of numerical regulators in the hydrodynamic simulations.

In this work, we will systematically examine the full non-linear causality conditions derived in Ref.~\cite{Bemfica:2020xym} in event-by-event simulations of relativistic heavy-ion collisions. We will further quantify the role of additional second-order gradient terms in the DNMR theory on flow observables. 

\section{The Hydrodynamic Framework}
\label{sec:model}

The hydrodynamic equations of motion for the collision system's stress-energy tensor represent the energy-momentum conservation as,
\begin{equation}
    \partial_\mu T^{\mu\nu} = 0 \,.
\label{eq:hydro}
\end{equation}
In the Israel-Stewart and the DNMR formalisms \cite{Israel:1976tn, Israel:1979wp, Muller:1967zza, Denicol:2012cn}, the out-of-equilibrium shear stress tensor and bulk viscous pressure are treated as independent degrees of freedom, and they evolve with the following relaxation-type of equations,
\begin{eqnarray}
	\tau_{\Pi} \dot{\Pi} + \Pi
	& = & - \zeta\, \theta
	- \delta_{\Pi \Pi} \Pi \, \theta
	+ \lambda_{\Pi \pi} \pi^{\mu\nu} \sigma_{\mu\nu} \label{eq:bulk} \\ 
	\tau_{\pi} \dot{\pi}^{\langle \mu\nu \rangle}
	+ \pi^{\mu\nu}
	& = & 2\eta\,\sigma^{\mu\nu}
	- \delta_{\pi \pi} \pi^{\mu\nu} \theta
	+ \varphi_7 \pi_{\alpha}^{\langle \mu} \pi^{\nu\rangle \alpha} \nonumber \\
	& & - \tau_{\pi \pi} \pi_{\alpha}^{\langle \mu} \sigma^{\nu\rangle \alpha}
	+ \lambda_{\pi \Pi} \Pi\, \sigma^{\mu\nu}\,.
	\label{eq:shear}
\end{eqnarray}
Here $A^{\langle \cdot  \cdot \rangle}$ denotes symmetrized and traceless projections, $\theta = \nabla_{\mu} u^{\mu}$ is the expansion rate, and 
\begin{equation}
\sigma^{\mu\nu}
= \frac{1}{2}\left[ \nabla^{\mu} u^{\nu} + \nabla^{\nu} u^{\mu}
- \frac{2}{3} \Delta^{\mu\nu} (\nabla_{\alpha} u^{\alpha}) \right]
\end{equation}
the velocity shear tensor, with $\nabla_{\mu} = ( g_{\mu\nu} - u_{\mu} u_{\nu} ) \partial^{\nu}$. The first-order transport coefficients $\eta$ and $\zeta$ are the shear and bulk viscosity, respectively. The numerical values of all the other second-order transport coefficients $\{\tau_\pi, \tau_\Pi, \delta_{\Pi\Pi}, \lambda_{\Pi\pi}, \delta_{\pi\pi}, \tau_{\pi\pi}, \lambda_{\pi\Pi}, \varphi_7\}$ are summarized in Table~\ref{table:transport_coeffs}.

The non-linear causality conditions for the DNMR theory were derived in Ref.~\cite{Bemfica:2020xym}. Here, we rewrite the causality inequality equations in terms of unitless ratios of the viscous pressure tensors over the enthalpy, $\varepsilon + P$. The non-zero eigenvalues of the shear stress tensor are denoted as $\{\Lambda_i\} (i = 1, 2, 3)$. The traceless condition of the shear stress tensor requires $\Lambda_1 + \Lambda_2 + \Lambda_3 = 0$. We also define the unitless coefficients $C_\eta = \tau_\pi(\varepsilon + P)/\eta$ and $C_\zeta = \tau_\Pi(\varepsilon + P)/\zeta$ for the ratios of the shear and bulk relaxation times to the shear and bulk viscosity, respectively. The shear coefficient $C_\eta$ is often approximated as a constant from $5 \sim 7$ in kinetic theories \cite{York:2008rr, Ghiglieri:2018dgf} or $(4 - 2\ln2)$ in the strongly-coupled theory \cite{Baier:2007ix, Finazzo:2014cna}. In this work, we set $C_\eta = 5$. The bulk coefficient $C_\zeta \propto 1/(1/3 - c_s^2)^\alpha$ with $\alpha = 1$ in the strongly-coupled theory \cite{Gubser:2008yx, Kanitscheider:2009as, Rougemont:2017tlu} and $\alpha = 2$ from the kinetic approach \cite{Huang:2010sa, Denicol:2012cn, Denicol:2014vaa}. In this work, we will examine hydrodynamic evolution with the following two choices for the bulk relaxation time,
\begin{equation}
    \tau_{\Pi, 1} = C_{\zeta, 1} \frac{\zeta}{\varepsilon + P} = \frac{1}{14.55(1/3 - c_s^2)^2} \frac{\zeta}{\varepsilon + P}
    \label{eq:tau_Pi1}
\end{equation}
and
\begin{equation}
    \tau_{\Pi, 2} = C_{\zeta, 2} \frac{\zeta}{\varepsilon + P} = \frac{C_\eta}{7 (1/3 - c_s^2)} \frac{\zeta}{\varepsilon + P}.
    \label{eq:tau_Pi2}
\end{equation}
The first choice $\tau_{\Pi, 1}$ was derived in kinetic theory \cite{Denicol:2014vaa} and was widely used in hydrodynamic simulations \cite{Ryu:2015vwa, Bernhard:2016tnd, Bernhard:2019bmu, Schenke:2020mbo, Everett:2020xug, Summerfield:2021oex}. The parametric form of the second choice $\tau_{\Pi, 2}$ is motivated from the strongly-coupled theory \cite{Gubser:2008yx, Kanitscheider:2009as, Huang:2010sa}.

\begin{figure}[ht!]
    \centering
    \includegraphics[width=0.9\linewidth]{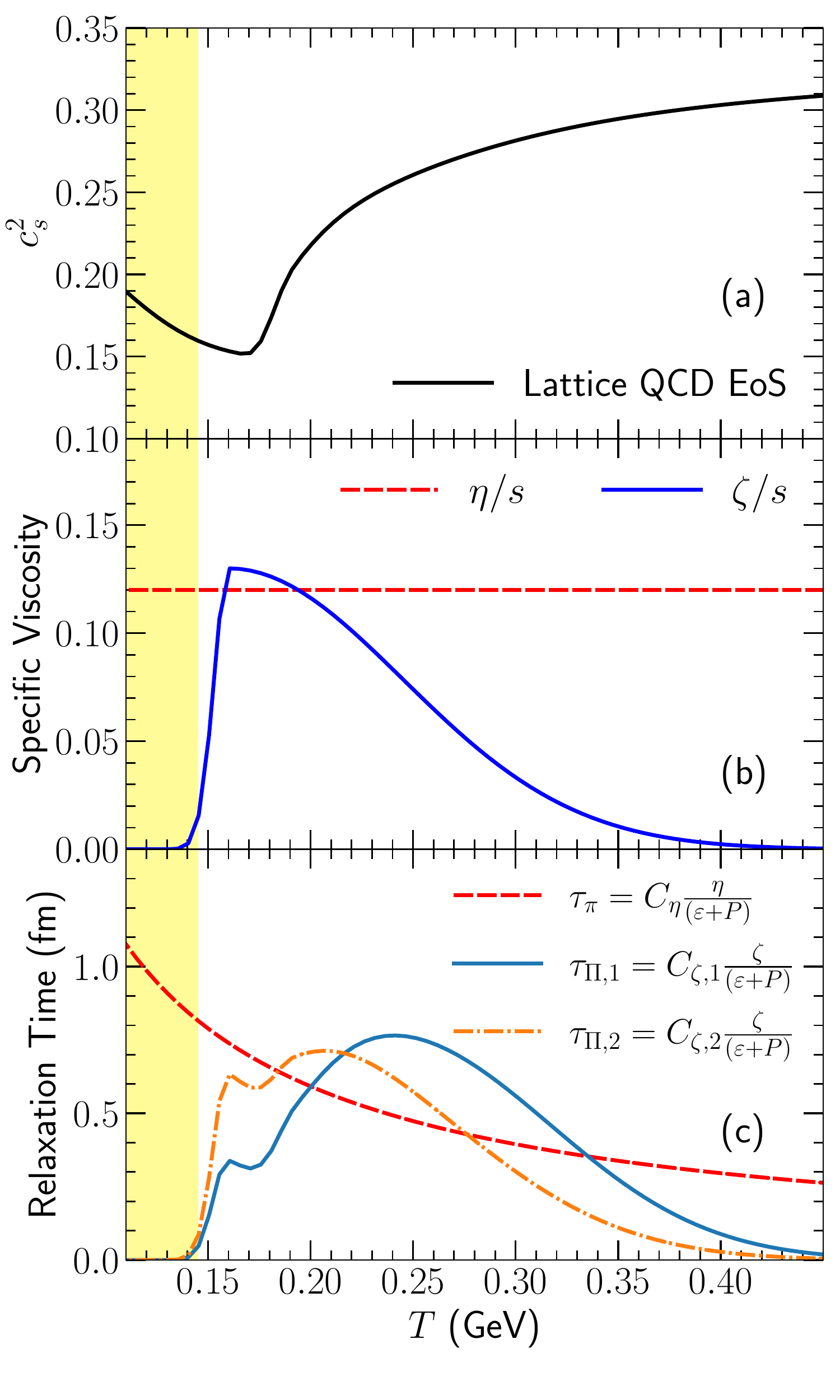}
  \caption{(Color online) Panel (a): The squared of the speed of sound from a Lattice QCD based EoS as a function of temperature. Panels (b) and (c): The specific shear and bulk viscosity and their relaxation times as functions of temperature used in this work. The region in yellow represents the phase ascribed to UrQMD in this work.}
  \label{fig:transportCoeff}
\end{figure}

According to Ref.~\cite{Bemfica:2020xym} the necessary conditions for causality can be written as,
\begin{equation}
    n_1 \equiv \frac{2}{C_\eta}+\frac{\lambda_{\pi\Pi}}{\tau_\pi}\frac{\Pi}{\varepsilon+P}-\frac{\tau_{\pi\pi}}{2\tau_\pi}\frac{|\Lambda_1|}{\varepsilon+P} \geq 0,
    \label{eq:causal_n1}
\end{equation}
\begin{equation}
    n_2 \equiv 1 - \frac{1}{C_\eta}+\left(1-\frac{\lambda_{\pi\Pi}}{2\tau_\pi}\right)\frac{\Pi}{\varepsilon+P}-\frac{\tau_{\pi\pi}}{4\tau_\pi}\frac{\Lambda_3}{\varepsilon+P} \geq 0,
    \label{eq:causal_n2}
\end{equation}
\begin{equation}
    n_3 \equiv \frac{1}{C_\eta}+\frac{\lambda_{\pi\Pi}}{2\tau_\pi}\frac{\Pi}{\varepsilon+P}-\frac{\tau_{\pi\pi}}{4\tau_\pi}\frac{\Lambda_3}{\varepsilon+P} \geq 0,
    \label{eq:causal_n3}
\end{equation}
\begin{eqnarray}
   \!\!\!\!\!\!\!\! n_4 &\equiv& 1 - \frac{1}{C_\eta} + \left(1-\frac{\lambda_{\pi\Pi}}{2\tau_\pi}\right)\frac{\Pi}{\varepsilon+P} \nonumber \\
    && + \left(1-\frac{\tau_{\pi\pi}}{4\tau_\pi} \right) \frac{\Lambda_a}{\varepsilon+P}-\frac{\tau_{\pi\pi}}{4\tau_\pi}\frac{\Lambda_d}{\varepsilon+P}\geq0,\, (a\neq d)
    \label{eq:causal_n4}
\end{eqnarray}
\begin{eqnarray}
    n_5 &\equiv& c^2_s + \frac{4}{3}\frac{1}{C_\eta} + \frac{1}{C_\zeta} + \left(\frac{2}{3}\frac{\lambda_{\pi\Pi}}{\tau_\pi} + \frac{\delta_{\Pi\Pi}}{\tau_\Pi} + c^2_s \right)\frac{\Pi}{\varepsilon+P} \nonumber \\
    && + \left( \frac{3\delta_{\pi\pi} + \tau_{\pi\pi}}{3\tau_\pi} + \frac{\lambda_{\Pi\pi}}{\tau_\Pi} + c^2_s\right) \frac{\Lambda_1}{\varepsilon + P} \geq 0,
    \label{eq:causal_n5}
\end{eqnarray}
\begin{eqnarray}
    \!\!\!\!\!\!\!\! n_6 &\equiv& 1 - \left(c^2_s + \frac{4}{3}\frac{1}{C_\eta} + \frac{1}{C_\zeta}\right) \nonumber \\
    && + \left(1 - \frac{2}{3}\frac{\lambda_{\pi\Pi}}{\tau_\pi} - \frac{\delta_{\Pi\Pi}}{\tau_\Pi} - c^2_s\right)\frac{\Pi}{\varepsilon+P} \nonumber \\
    && + \left(1 - \frac{3\delta_{\pi\pi} + \tau_{\pi\pi}}{3\tau_\pi} - \frac{\lambda_{\Pi\pi}}{\tau_\Pi} - c^2_s\right)\frac{\Lambda_3}{\varepsilon + P} \geq 0.
    \label{eq:causal_n6}
\end{eqnarray}
The original necessary conditions $n_5$ and $n_6$ in Ref.~\cite{Bemfica:2020xym} are simplified with the condition, $\left( \frac{\delta_{\pi\pi}}{\tau_\pi} + \frac{\tau_{\pi\pi}}{3\tau_\pi} + \frac{\lambda_{\Pi\pi}}{\tau_\Pi} + c^2_s\right) > 1$, in Eqs.~(\ref{eq:causal_n5}) and (\ref{eq:causal_n6}) according to the values of second-order transport coefficients in Table~\ref{table:transport_coeffs}~\cite{Denicol:2014vaa}.
%
\begin{figure}[b!]
    \centering
    \includegraphics[width=1.0\linewidth]{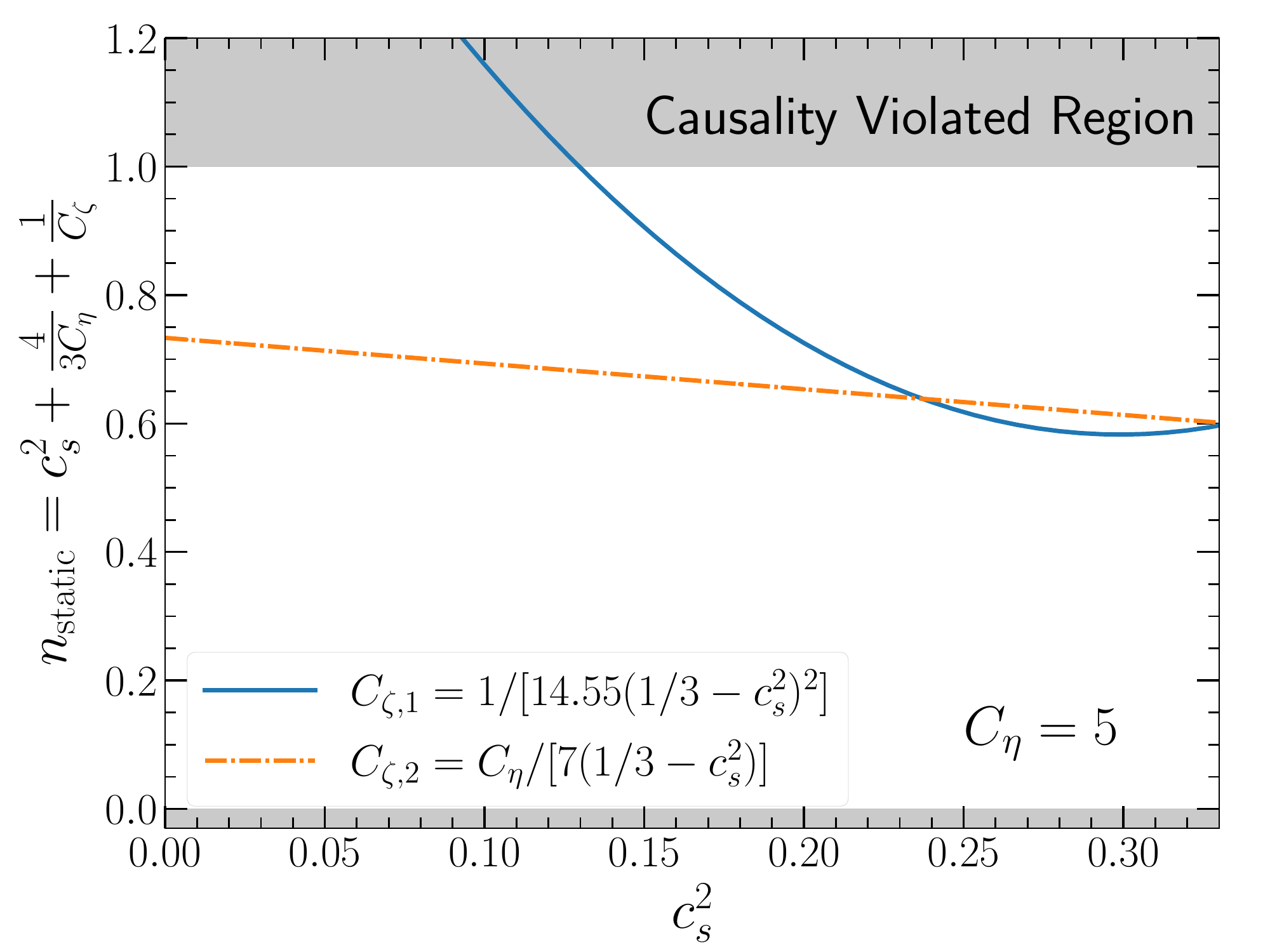}
  \caption{(Color online) The causal region for two choices of relaxation time of bulk viscosity as a function of the square of speed of sound $c_s^2$ in the absence of viscous corrections, $\Pi = \Lambda_i = 0$ $(i = 1, 2, 3)$.}
  \label{fig:staticCausalRegion}
\end{figure}
%
The sufficient causality conditions can be found in the Appendix. We first examine the causality constraints with different functional forms of $C_\zeta$ in the absence of viscous corrections, at which $\Pi = \Lambda_i = 0$ $(i = 1, 2, 3)$. In this case, both the necessary and sufficient causality conditions reduce to
\begin{equation}
    0 \le n_\mathrm{static} \equiv c_s^2 + \frac{4}{3C_\eta} + \frac{1}{C_\zeta} \le 1.
\end{equation}
Figure~\ref{fig:staticCausalRegion} shows the value of $n_\mathrm{static}$ as a function of the speed of sound squared for the two choices of $C_\zeta$. At the conformal limit $c_s^2$ goes to $1/3$, $C_\zeta$ approaches $+\infty$ and $n_\mathrm{static}$ approaches $(C_\eta + 4)/(3 C_\eta) = 0.6$ when $C_\eta = 5$. The $C_{\zeta, 1}$ from the kinetic theory with relaxation time approximation \cite{Denicol:2014vaa} has a quadratic dependence on $c_s^2$, which leads to a rapid increase of $n_\mathrm{static}$ at small $c_s^2$ values. With the coefficient $14.55$ in $C_{\zeta, 1}$, the $n_\mathrm{static}$ exceeds the causality bound for $c_s^2 < 0.13$. Although the minimum $c_s^2$ in the lattice QCD EoS at zero net baryon density is around 0.15 (shown in Fig.~\ref{fig:transportCoeff}a below), this choice of $C_{\zeta, 1}$ leads to a strong restriction on the sizes of the viscous stress tensor $\pi^{\mu\nu}$ and $\Pi$ near the smooth crossover region where $c_s^2 \sim 0.15$. If the EoS has a soft point $c_s^2 \rightarrow 0$ at some finite baryon density, $C_{\zeta, 1}$ will be ruled out by the causality conditions. To ensure $0 \le n_\mathrm{static} \le 1$ for $0 \le c_s^2 \le 1/3$, the quadratic $c_s^2$ parameterization requires the coefficient to be less than $9 - (12/C_\eta) = 6.6$ for $C_\eta = 5$. On the other hand, the strongly-coupled theory suggests a linear dependence of $c_s^2$ in $1/C_\zeta$. Figure~\ref{fig:staticCausalRegion} shows that $n_\mathrm{static}$ with $C_{\zeta,2}$ increases much slower than that with $C_{\zeta, 1}$.

%
%
Figure~\ref{fig:transportCoeff} shows the speed of sound squared from a lattice QCD EoS \cite{Bazavov:2014pvz, Moreland:2015dvc} and the transport coefficients that will be used to simulate relativistic heavy-ion collisions. We use the specific shear and bulk viscosity from Ref.~\cite{Schenke:2020mbo}.
The two choices of the bulk relaxation time $\tau_\Pi$ are shown in Fig.~\ref{fig:transportCoeff}c. We will examine event-by-event hydrodynamic simulations for Au+Au collisions and p+Au collisions at the top RHIC energy. We expect a longer hydrodynamic phase at higher LHC energies, which will reduce the flow observables' sensitivity to the choice of the bulk relaxation time compared to those at the RHIC. 

\begin{table*}
\begin{tabular}{|c|c|c|c|c|}
\hline
     & restricted DNMR with $\tau_{\Pi, 1}$ & restricted DNMR with $\tau_{\Pi, 2}$ & DNMR with $\tau_{\Pi, 1}$  & DNMR with $\tau_{\Pi, 2}$ \\ \hline
  $\frac{\eta}{\tau_\pi (\varepsilon + P)}=\frac{1}{C_\eta}$   & $\frac{1}{5}$ & $\frac{1}{5}$ & $\frac{1}{5}$ & $\frac{1}{5}$ \\ \hline
  $\frac{\zeta}{\tau_\Pi (\varepsilon + P)}=\frac{1}{C_\zeta}$   &  $14.55(\frac{1}{3}-c_s^2)^2$ & $\frac{7}{C_\eta} (\frac{1}{3} - c_s^2)$ &  $14.55(\frac{1}{3}-c_s^2)^2$ & $\frac{7}{C_\eta} (\frac{1}{3} - c_s^2)$ \\ \hline
  $\frac{\delta_{\pi\pi}}{\tau_\pi}$   & $\frac{4}{3}$ & $\frac{4}{3}$ & $\frac{4}{3}$ & $\frac{4}{3}$ \\ \hline
  $\frac{\delta_{\Pi\Pi}}{\tau_\Pi}$   & $\frac{2}{3}$ & $\frac{2}{3}$ & $\frac{2}{3}$ & $\frac{2}{3}$ \\ \hline
  $\frac{\tau_{\pi\pi}}{\tau_\pi}$   & $0$ & $0$ & $\frac{10}{7}$ & $\frac{10}{7}$ \\ \hline
  $\frac{\lambda_{\pi\Pi}}{\tau_\pi}$   & $0$ & $0$ & $\frac{6}{5}$ & $\frac{6}{5}$ \\ \hline
  $\frac{\lambda_{\Pi\pi}}{\tau_\Pi}$   & $0$ & $0$ & $\frac{8}{5}(\frac{1}{3} - c_s^2)$ & $\frac{8}{5}(\frac{1}{3} - c_s^2)$ \\ \hline
  $\varphi_7$ & $0$ & $0$ & $\frac{9}{70} \frac{4}{\varepsilon + P}$ & $\frac{9}{70} \frac{4}{\varepsilon + P}$ \\ \hline
\end{tabular}
\caption{The choice of second-order transport coefficients used in the restricted and full DNMR hydrodynamic theories. \cite{Denicol:2014vaa, Kanitscheider:2009as}}
\label{table:transport_coeffs}
\end{table*}

Table~\ref{table:transport_coeffs} summarizes the numerical values of all the second-order transport coefficients that we will use in this work. We refer to the simulations with non-zero values $\{\tau_{\pi\pi}, \lambda_{\pi\Pi}, \lambda_{\Pi \pi}, \varphi_7\}$ as the DNMR theory, while the restricted DNMR theory, which is close to the original Israel-Stewart hydrodynamics, only includes the relaxation times for shear and bulk viscosity and $\delta_{\pi\pi}$ and $\delta_{\Pi\Pi}$ terms in the equations of motion.

In hydrodynamic simulations, it is practical to track the evolution of the inverse Reynolds numbers for the shear stress tensor,
\begin{equation}
    R_\pi \equiv \frac{\sqrt{\pi^{\mu\nu} \pi_{\mu\nu}}}{\varepsilon + P}
    \label{eq:Rpi}
\end{equation}
and for the bulk viscous pressure,
\begin{equation}
    R_\Pi \equiv \frac{\Pi}{\varepsilon + P}.
\end{equation}
From the similarity transformation, the inverse Reynolds number $R_\pi$ is related to the eigenvalues of the shear stress tensor by
\begin{equation}
    R_\pi = \sqrt{\left(\frac{\Lambda_1}{\varepsilon+P}\right)^2 +\left(\frac{\Lambda_2}{\varepsilon+P}\right)^2 +\left(\frac{\Lambda_3}{\varepsilon+P}\right)^2}.
\end{equation}
Using the tracelessness condition for the shear stress tensor, we can derive the following inequality,
\begin{equation}
   \frac{\sqrt{6}}{2}\frac{\Lambda_\mathrm{max}}{\varepsilon + P} \le R_\pi \le \sqrt{2} \frac{\Lambda_\mathrm{max}}{\varepsilon + P},
   \label{eq:RpiSize}
\end{equation}
where $\Lambda_\mathrm{max} \equiv \mathrm{max}\{\vert \Lambda_1 \vert, \vert \Lambda_2 \vert, \vert \Lambda_3 \vert\}$ is the maximum absolute eigenvalue of $\pi^{\mu\nu}$.

\section{Results}\label{sec:results}

\subsection{Visualizing the causal regions with inverse Reynolds numbers}

Before examining the causality conditions in event-by-event hydrodynamic simulations, we first identify the causal region in terms of the inverse Reynolds numbers.

\begin{figure*}[ht!]
  \centering
  \begin{tabular}{cc}
    \includegraphics[width=0.45\linewidth]{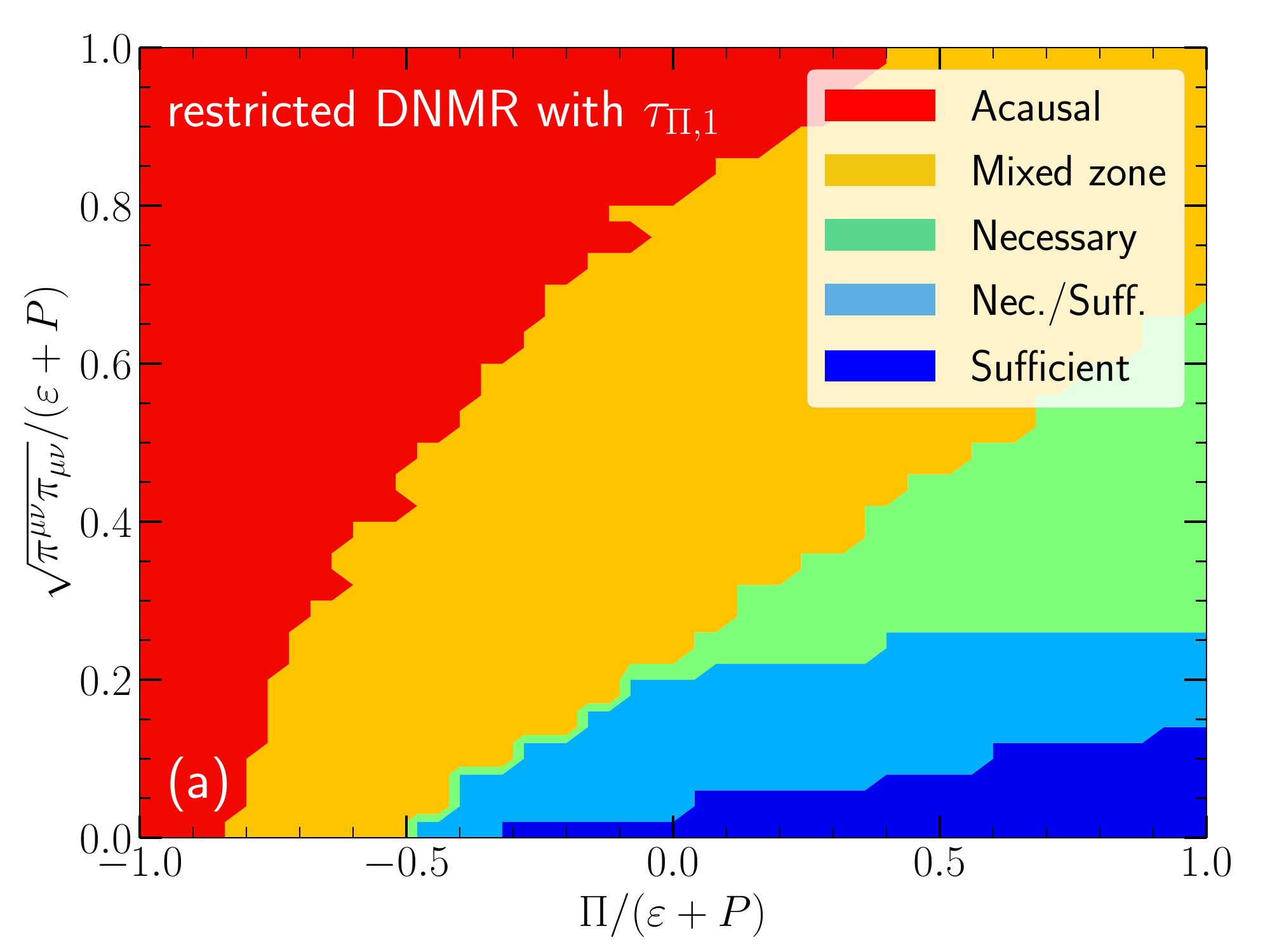} &  \includegraphics[width=0.45\linewidth]{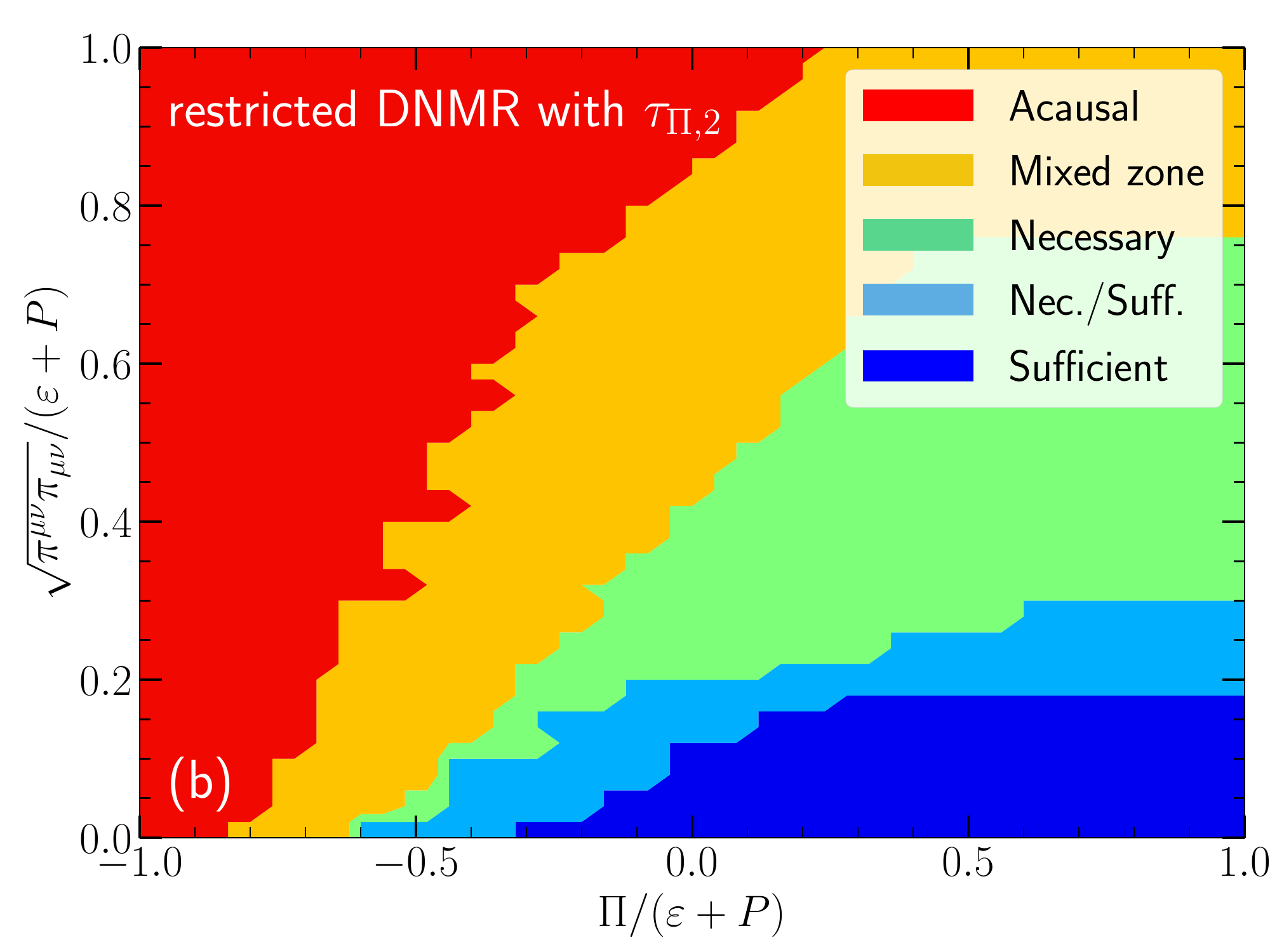} \\
      \includegraphics[width=0.45\linewidth]{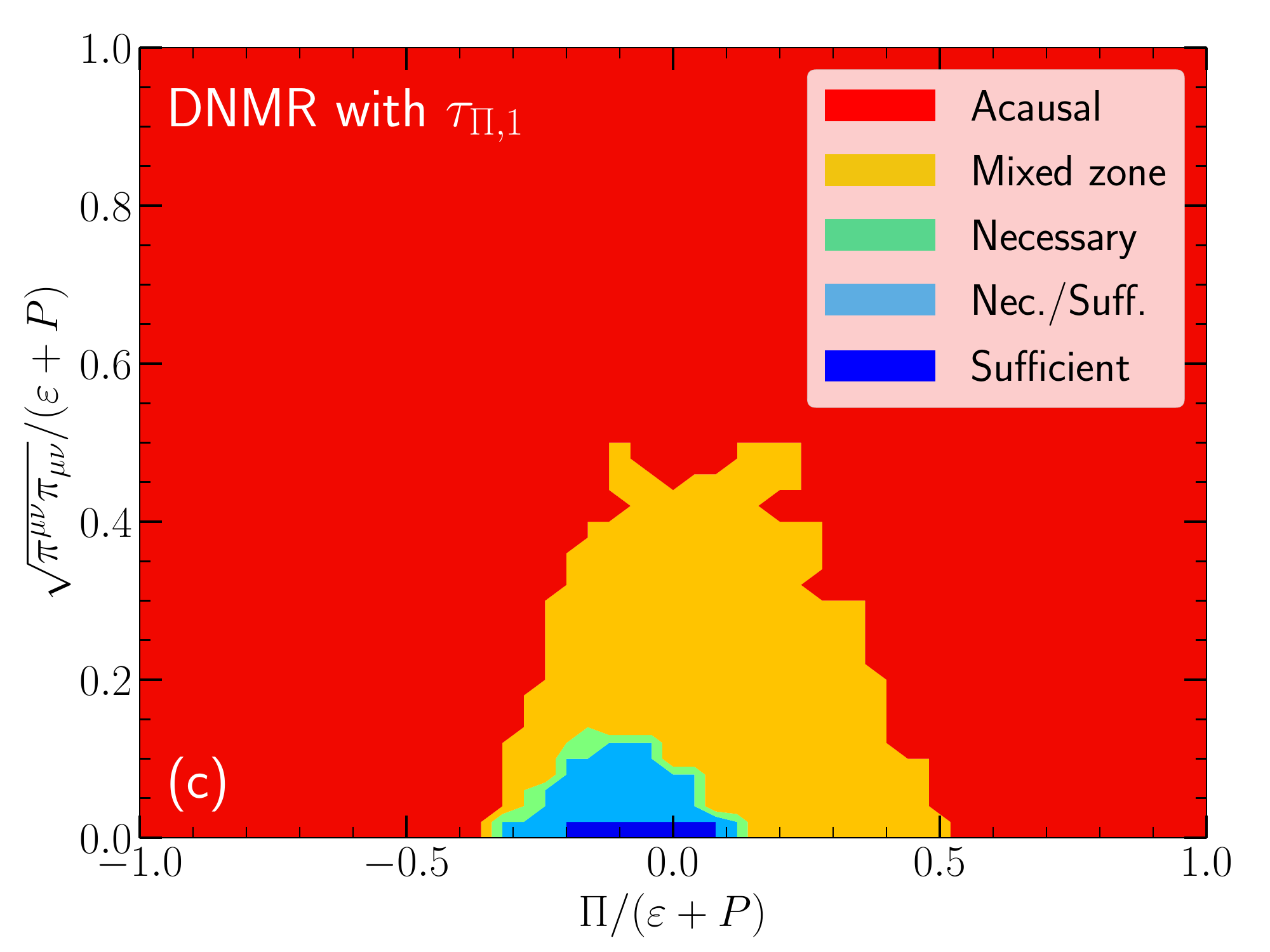} &  \includegraphics[width=0.45\linewidth]{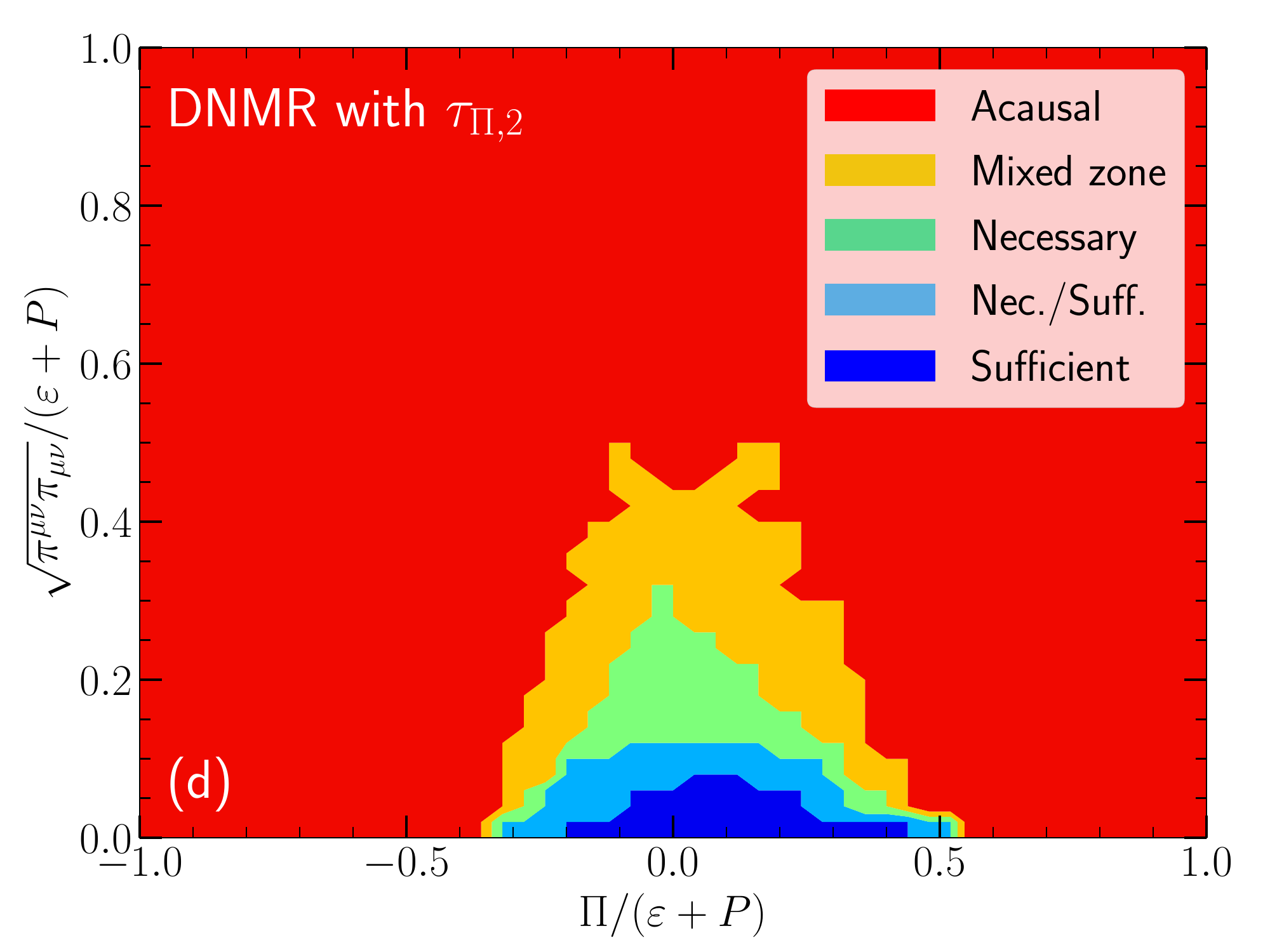}
  \end{tabular}
  \caption{(Color online) Causal regions as functions of the inverse Reynolds numbers for the shear and bulk viscosity. The restricted and full DNMR hydrodynamic theories with two different forms of the bulk relaxation time are presented.}
  \label{fig:causalregions}
\end{figure*}
%
Figure~\ref{fig:causalregions} demonstrates the causal regions with different choices of second-order transport coefficients listed in Table~\ref{table:transport_coeffs}. We test both the necessary and sufficient causality conditions in a 5-dimensional space of $\{c_s^2, \Lambda_1, \Lambda_2, \Lambda_3, \Pi\}$, where $c_s^2$ varies between 0.15 to 1/3 and $\{\Lambda_i\}$ and $\Pi$ varies from 0 to $\varepsilon + P$. Here, we present the causal regions in terms of the shear and bulk inverse Reynolds numbers. These two variables represent how far a fluid cell is out-of-equilibrium at a given space-time position. The red regions in the Fig.~\ref{fig:causalregions} violate the necessary causality conditions. Fluid cells in this region violate causality for sure. The yellow bands indicate a mixed region, where some fluid cells satisfy the necessary causality conditions and some are acasual, depending on exact values of $c_s^2$ and the shear $\{\Lambda_i\}$. Fluid cells with $R_\pi$ and $R_\Pi$ in the green areas satisfy the necessary causality conditions but violate the sufficient conditions. The light blue region contains a mixture of fluid cells that satisfy or violate the sufficient conditions. The current causality conditions are not sufficient to determine whether fluid cells are causal or not in yellow, green, and light blue regions. Finally, the dark blue regions show the inverse Reynolds numbers allowed by the sufficient causality conditions in which fluid cells are causal.

The causality conditions impose maximum allowed values for the inverse Reynolds numbers $R_\pi$ and $\vert R_\Pi \vert$ during the hydrodynamic evolution. Because the bulk viscous pressure acts against the local expansion rate $\theta = \partial_\mu u^\mu$, its value relaxes to the negative Navier-Stokes value $\Pi \sim - \zeta \theta$. So $R_{\Pi}$ is negative in most cases.
Figures~\ref{fig:causalregions} show that the necessary causality conditions require $R_\pi < 1$ and $\vert R_\Pi \vert < 1$ when $R_\Pi < 0$ for the DNMR hydrodynamics. The sufficient conditions require extremely small viscous corrections, $R_\pi \le 0.1$ when the bulk viscous pressure is negative.

Comparing Fig.~\ref{fig:causalregions}a with \ref{fig:causalregions}b, we find that the bulk relaxation time $\tau_{\Pi, 2}$ allows a larger causal region than that with $\tau_{\Pi, 1}$, which is consistent with the static conditions $n_\mathrm{static}$ shown in Fig.~\ref{fig:staticCausalRegion}. The mixed zone is large for $\tau_{\Pi, 1}$ because of its fast quadratic dependence on $c_s^2$. For Israel-Stewart theory, the causal region becomes bigger when $R_\Pi > 0$, when the terms with $R_\Pi$ give opposite contributions in Eqs.~(\ref{eq:causal_n1})-(\ref{eq:causal_n6}) compared to those with the shear stress tensor. 

Comparing Fig.~\ref{fig:causalregions}a(b) with \ref{fig:causalregions}c(d), we find the allowed causal region shrinks significantly, especially for $R_\Pi > 0$ for the full DNMR theory. Because the sign for $\left(1 - \frac{2}{3}\frac{\lambda_{\pi\Pi}}{\tau_\pi} - \frac{\delta_{\Pi\Pi}}{\tau_\Pi} - c^2_s\right)$ in Eq.~(\ref{eq:causal_n6}) flips from positive to negative with $\lambda_{\pi\Pi}/\tau_\pi = 6/5$, regions with large positive $R_\Pi$ values are not allowed anymore.
Figures \ref{fig:causalregions}a(b) vs. \ref{fig:causalregions}c(d) visually show that the non-zero second-order transport coefficients $\{\tau_{\pi\pi}, \lambda_{\pi\Pi}, \lambda_{\Pi\pi}\}$ set strong restrictions on the size of shear stress tensor $\pi^{\mu\nu}$ and bulk viscous pressure $\Pi$.
With these additional second-order transport coefficients, the inverse Reynolds numbers need to be smaller than 0.5 for both choices of bulk relaxation time.

In practice, numerical simulations were found to be stable when $R_\pi \le 1$ and $\vert R_\Pi \vert \le 1$ with a grid spacing $dx = 0.067$\,fm \cite{Schenke:2020mbo}. These stability conditions are less demanding compared to the causal regions shown in Fig.~\ref{fig:causalregions}.

\subsection{Examining realistic hydrodynamic simulations}

After identifying the causal regions in terms of inverse Reynolds number, we would like to find the limits on $R_\pi$ and $R_\Pi$ to ensure all fluid cells in hydrodynamic simulations are within the causal region. Those limits would be handy when performing large-scale event-by-event simulations.

\begin{figure}[t!]
  \centering
    \includegraphics[width=\linewidth]{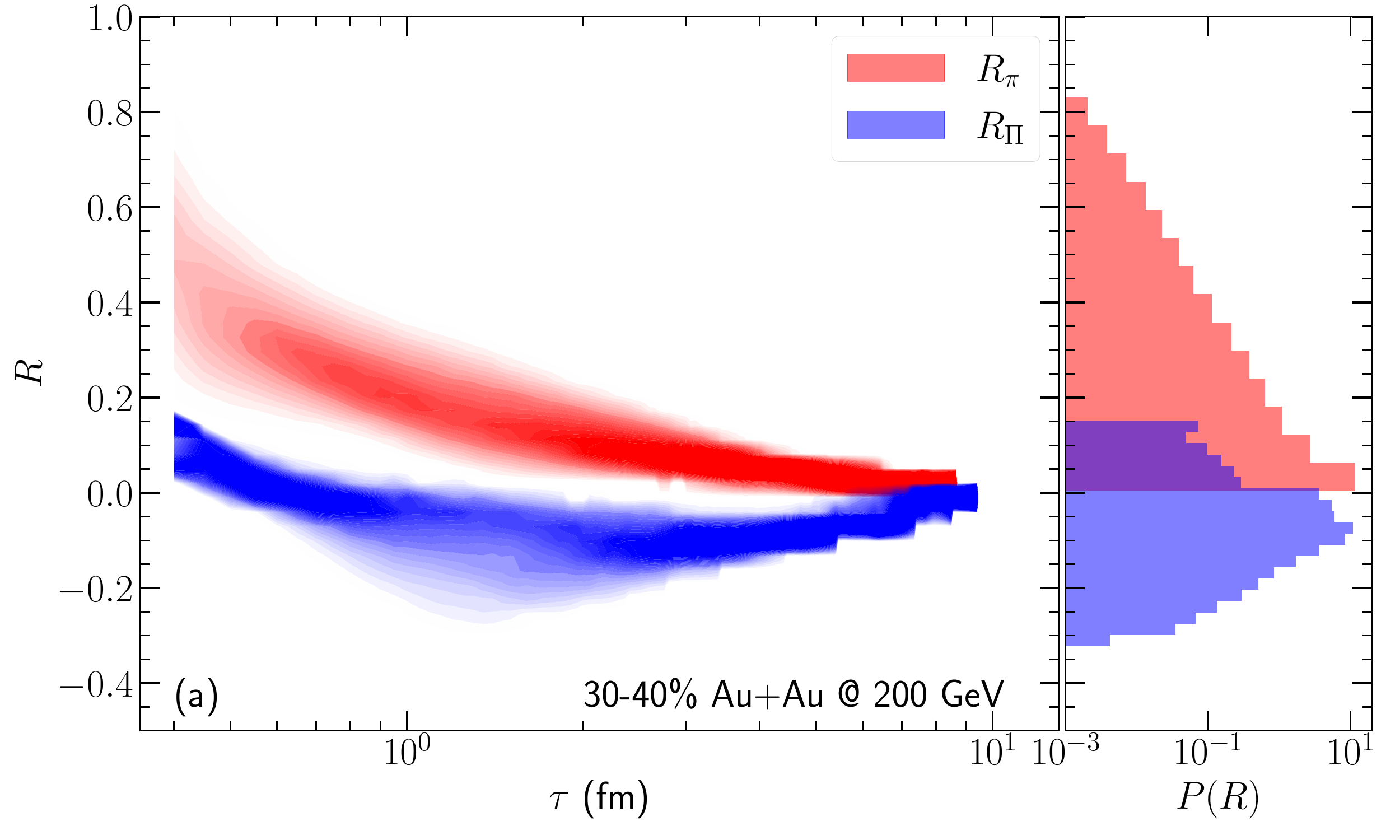}
    \includegraphics[width=\linewidth]{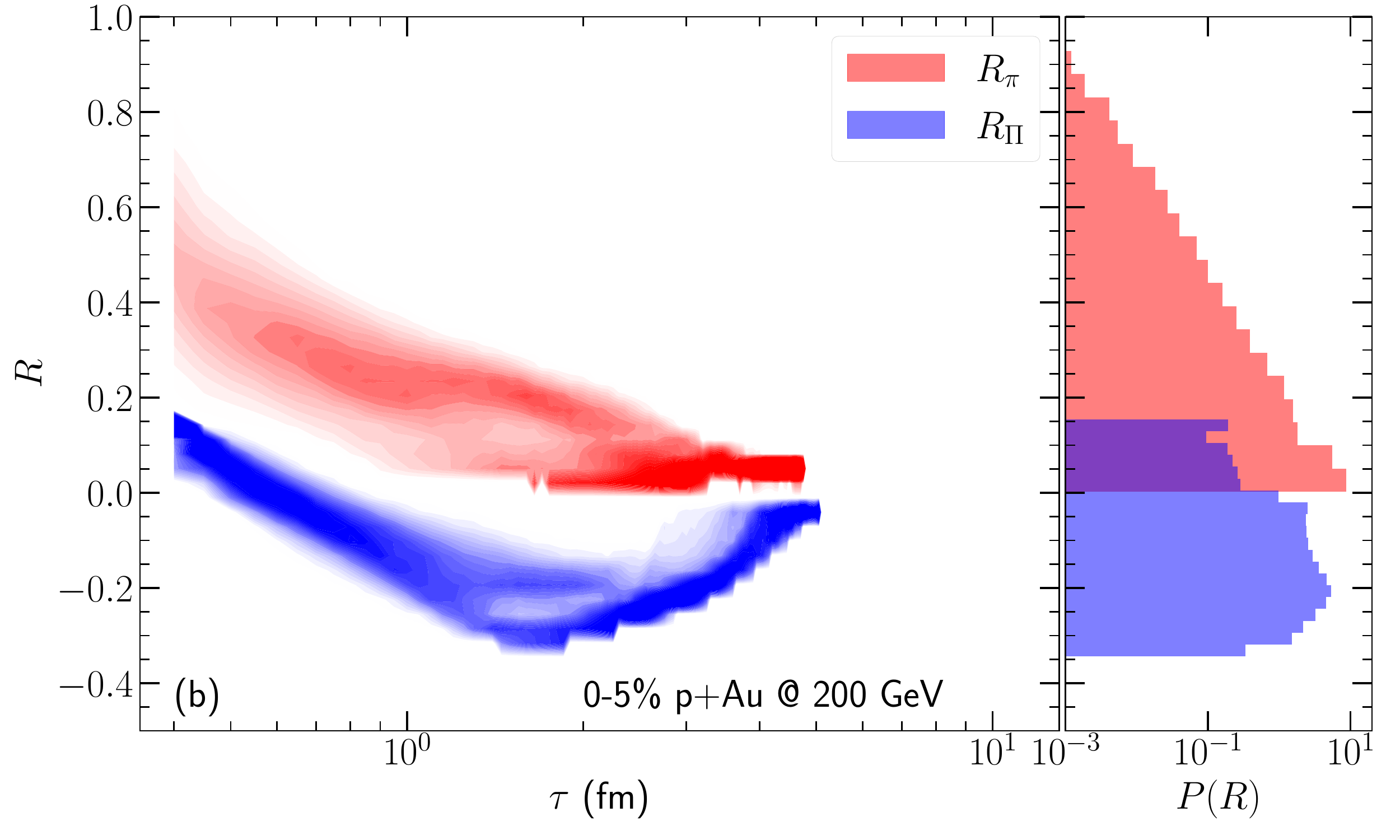}
  \caption{(Color online) The distributions of fluid cells' inverse Reynolds numbers as a function of the longitudinal proper time for one 30-40\% Au+Au collision (a) and one 0-5\% p+Au collision (b) at the top RHIC energy. The right panels show the time-integrated probability distributions of the inverse Reynolds numbers.}
  \label{fig:InvReynoldsNumDis}
\end{figure}

It is instructive to first study the inverse Reynolds numbers' distributions as functions of the evolution time in relativistic hydrodynamic simulations. We analyze typical hydrodynamic evolution for 30-40\% Au+Au and 0-5\% p+Au collisions at the top RHIC energy using the IP-Glasma + MUSIC + UrQMD framework \cite{Schenke:2020mbo}. Figure~\ref{fig:InvReynoldsNumDis} shows the time-dependent and time-integrated distributions of the $R_\pi$ and $R_\Pi$ for fluid cells with temperature $T \ge 145$\,MeV.

The maximum of the shear inverse Reynolds number reaches around one at the starting time of hydrodynamics $\tau_0 = 0.4$\,fm/$c$. Most of the fluid cells have an average $R_\pi \sim 0.3$ during the first fm/$c$ of the hydrodynamic evolution in 30-40\% Au+Au and 0-5\% p+Au collisions. The bulk inverse Reynolds numbers start with positive values $(0.05-0.15)$ at $\tau_0 = 0.4$ fm/$c$ to compensate the difference between the trace anomaly in lattice QCD EoS and the traceless energy stress tensor from the IP-Glasma phase \cite{Mantysaari:2017cni, Schenke:2020mbo}. As the bulk viscous pressure evolves towards its Navier-Stokes limit $-\zeta \theta$, most of the fluid cells' $R_\Pi$ evolve to negative values during the first 0.5 fm/$c$ and reach their minima around $\tau = 1.5-2$ fm/$c$. Since the more compact 0-5\% p+Au collision develops a larger expansion rate than that in 30-40\% Au+Au collisions, the $R_\Pi$'s distribution has a peak around $-0.2$ in central p+Au collisions while most of the fluid cells have $\vert R_\Pi \vert < 0.1$ in 30-40\% Au+Au collisions. 
For $\tau > 2$ fm/$c$, the absolute values of $R_\pi$ and $R_\Pi$ decreases with $\tau$ because the local velocity gradients decrease with the evolution time.

\begin{figure}[t!]
  \centering
    \includegraphics[width=1.0\linewidth]{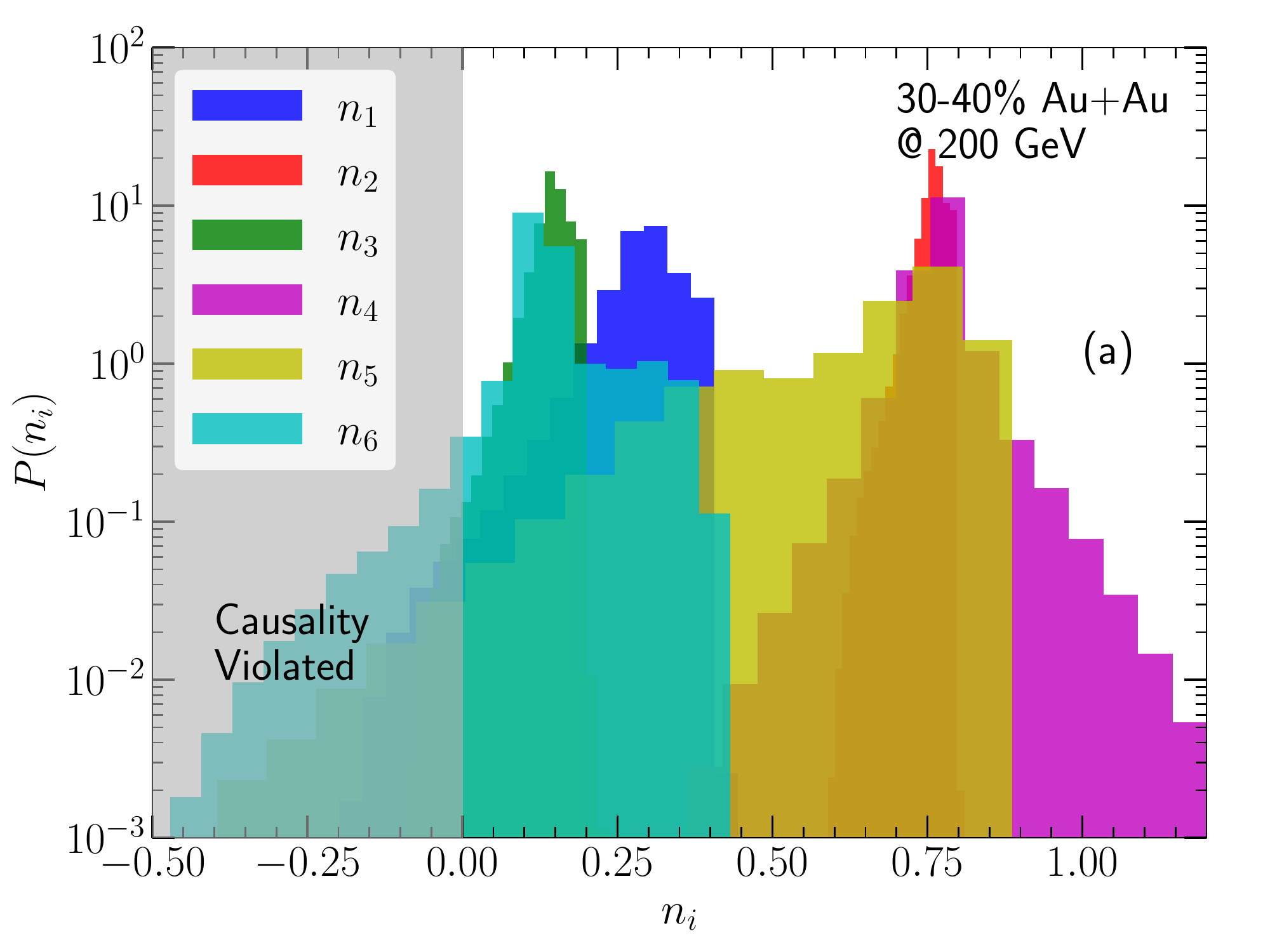}
    \includegraphics[width=1.0\linewidth]{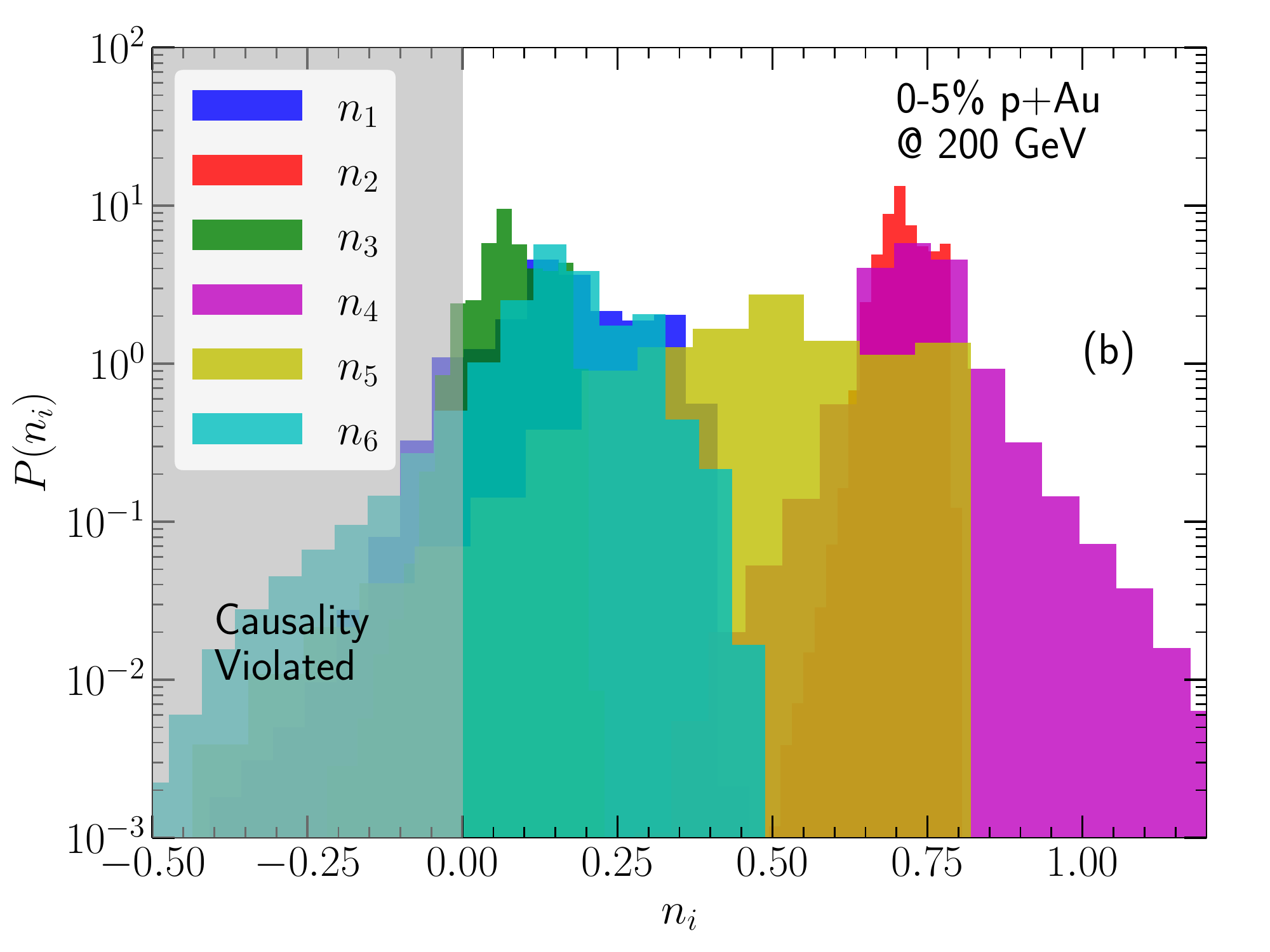}
  \caption{(Color online) Probability distributions for the necessary causality measures in fluid cells with temperature above 145 MeV in a typical Au+Au collision at 30-40\% centrality (a) and a 0-5\% p+Au collision (b) at 200 GeV. Hydrodynamic evolution is simulated with the DNMR equation of motion and the bulk relaxation time $\tau_{\Pi, 1}$. During the evolution, we restrict $R_\pi \le 1$ and $\vert R_\Pi \vert \le 1$.}
  \label{fig:Pn_DNMR_tauPi1}
\end{figure}
\begin{figure}[t!]
  \centering
    \includegraphics[width=1.0\linewidth]{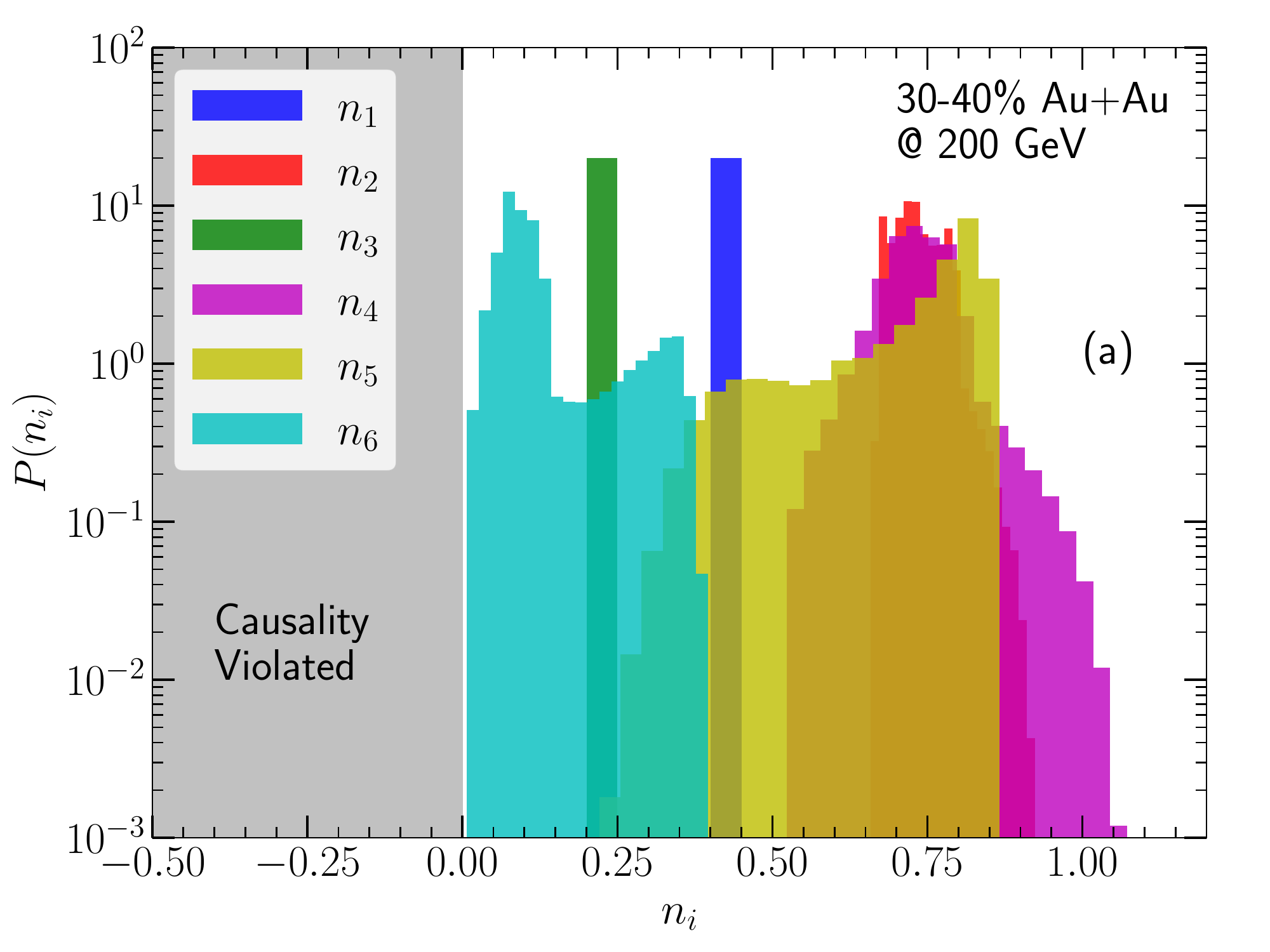}
    \includegraphics[width=1.0\linewidth]{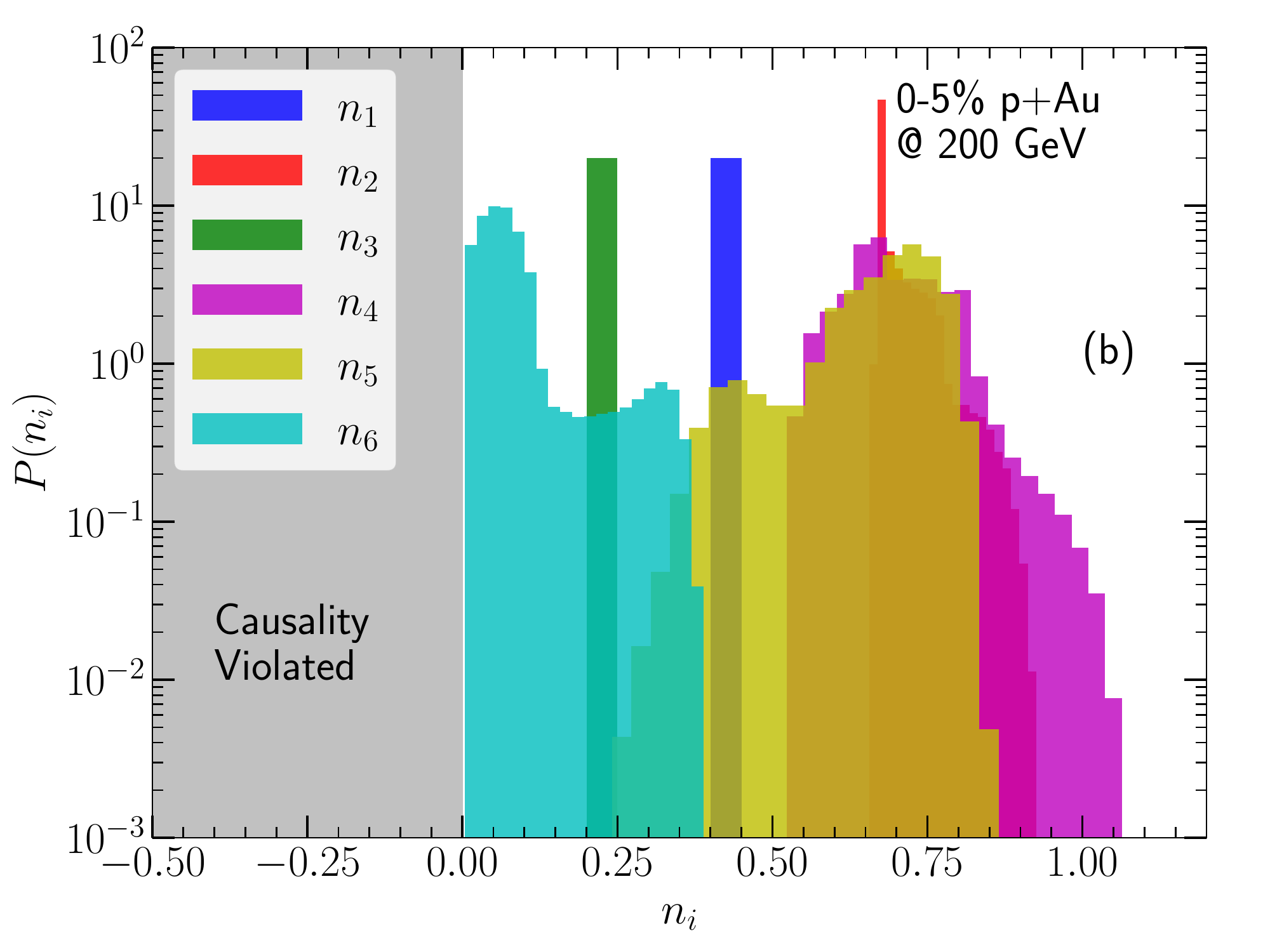}
  \caption{(Color online) Similar probability distributions as those in Fig.~\ref{fig:Pn_DNMR_tauPi1} but for hydrodynamic evolution simulated with the restricted DNMR equation of motion and the bulk relaxation time $\tau_{\Pi, 1}$. During the evolution, we restrict $R_\pi \le \sqrt{2} P/(\varepsilon+P)$ and $\vert R_\Pi \vert \le P/(\varepsilon+P)$.}
  \label{fig:Pn_IS_tauPi1}
\end{figure}

Figure~\ref{fig:Pn_DNMR_tauPi1} shows the probability distribution of each causality measure in Eqs.~(\ref{eq:causal_n1}) to (\ref{eq:causal_n6}) in realistic hydrodynamic simulations \cite{Schenke:2020mbo}. We find that there are 3.8\% of the fluid cells in 30-40\% Au+Au collisions violate the necessary causality conditions. The fraction of violating fluid cells increases to 17\% in 0-5\% p+Au collisions. The strong pressure gradients in the p+Au collision lead to fast expansion and drive the system out-of-equilibrium. Hence, simulating small systems is more challenging than large collision systems as they can evolve further away from local thermal equilibrium.

We find that the necessary causality conditions $n_1$, $n_3$, $n_5$, and $n_6$ effectively constrain the sizes of inverse Reynolds numbers in the DNMR hydrodynamics. Among them, the condition $n_6$ in Eq.~(\ref{eq:causal_n6}) imposes the strongest constraint. Table~\ref{tab:ViolationFraction} summarizes the fractions of fluid cells that violate necessary or sufficient causality conditions for different transport coefficient choices in hydrodynamic simulations. We note that there are significant fractions of fluid cells that satisfy the necessary conditions but violate the sufficient causality conditions. The current causality conditions can not determine whether they violate causality or not. As shown in Figs.~\ref{fig:causalregions}, we would need to impose very strong constraints $R_\pi \ll 1$ and $\vert R_\Pi \vert \ll 1$ to ensure all fluid cells to satisfy the sufficient conditions, which significantly limits simulating viscous effects in relativistic hydrodynamic evolution.
We note that Figure~\ref{fig:Pn_DNMR_tauPi1} and Table~\ref{tab:ViolationFraction} here summarize the overall fraction of fluid cells violate the necessary and sufficient causality conditions. Because the inverse Reynolds numbers are large during the early time of the evolution as shown in Figure~\ref{fig:InvReynoldsNumDis}, the violation of causality could potentially play a substantial role during the first few fm/$c$ of the hydrodynamic evolution. A quantitative time differential analysis was presented in a recent work~\cite{Plumberg:2021bme}.

\begin{table*}[ht!]
    \centering
    \begin{tabular}{|c|c|c|c|}
        \hline 
       Collision system & Transport coefficients &  Violate necessary conditions & Violate sufficient conditions \\ \hline 
        \multirow{2}{*}{30-40\% AuAu} & restricted DNMR with  $\tau_{\Pi, 1}$  & 1.8\% & 33\% \\ \cline{2-4}
            & DNMR with  $\tau_{\Pi, 1}$  & 3.8\% & 22\% \\ \hline
        \multirow{2}{*}{0-5\% pAu} & restricted DNMR with  $\tau_{\Pi, 1}$  & 9\% & 66\% \\ \cline{2-4}
            & DNMR with  $\tau_{\Pi, 1}$ & 17\% & 48\% \\ \hline
        \multirow{2}{*}{30-40\% AuAu} & restricted DNMR with $\tau_{\Pi, 2}$ & 0.1\% & 14\% \\ \cline{2-4}
            & DNMR with  $\tau_{\Pi, 2}$ & 1.7\% & 16\% \\ \hline
        \multirow{2}{*}{0-5\% pAu} & restricted DNMR with $\tau_{\Pi, 2}$ & 0.2\% & 25\% \\ \cline{2-4}
            & DNMR with $\tau_{\Pi, 2}$ & 7\% & 40\% \\ \hline
    \end{tabular}
    \caption{The fractions of cells violate necessary or sufficient causality conditions in 30-40\% Au+Au and 0-5\% p+Au collisions at 200 GeV with different choices of transport coefficients. We restrict the inverse Reynolds numbers $R_\pi \le 1$ and $\vert R_\Pi \vert \le 1$.}
    \label{tab:ViolationFraction}
\end{table*}

To regulate all fluid cells that violate the necessary causality conditions, we need to impose restrictions on inverse Reynolds numbers' sizes during hydrodynamic evolution. In Fig.~\ref{fig:Pn_IS_tauPi1}, we find that imposing $R_\pi \le \sqrt{2}P/(\varepsilon + P)$ and $\vert R_\Pi \vert \le P/(\varepsilon + P)$ can ensure all the fluid cells in the 30-40\% Au+Au and 0-5\% p+Au collisions at 200 GeV stay within the necessary causal region for the restricted DNMR hydrodynamics with $\tau_{\Pi, 1}$. By setting the transport coefficients $\lambda_{\pi \Pi}$ and $\tau_{\pi\pi}$ to zero, the necessary condition measures $n_1$ and $n_3$ reduce to static inequalities that only depend on $C_\eta$'s value. The condition $n_6$ imposes the dominant constraints.
In Eq.~(\ref{eq:RpiSize}), the inverse Reynolds number of the shear stress tenor $R_\pi \le \sqrt{2} \Lambda_\mathrm{max}/(\varepsilon + P)$. If we choose $\Lambda_\mathrm{max} = P$, then $R_\pi \le \sqrt{2}P/(\varepsilon + P)$.
Condition $\vert R_\Pi \vert \le P/(\varepsilon + P)$ is equivalent to $\vert \Pi \vert /P \le 1$, making sure that the thermal pressure is larger than the bulk viscous pressure and the total pressure is positive. Therefore, this condition avoids the formation of unstable cavitation regions during the evolution \cite{Torrieri:2008ip, Rajagopal:2009yw, Denicol:2015bpa, Byres:2019xld}.

We further examine the sufficient conditions after imposing the restrictions on the inverse Reynolds numbers and find the fractions of violating fluid cells remain almost unchanged as those in Table.~\ref{tab:ViolationFraction}. The detailed analysis is represented in the Appendix.

\begin{figure}[t!]
  \centering
    \includegraphics[width=0.9\linewidth]{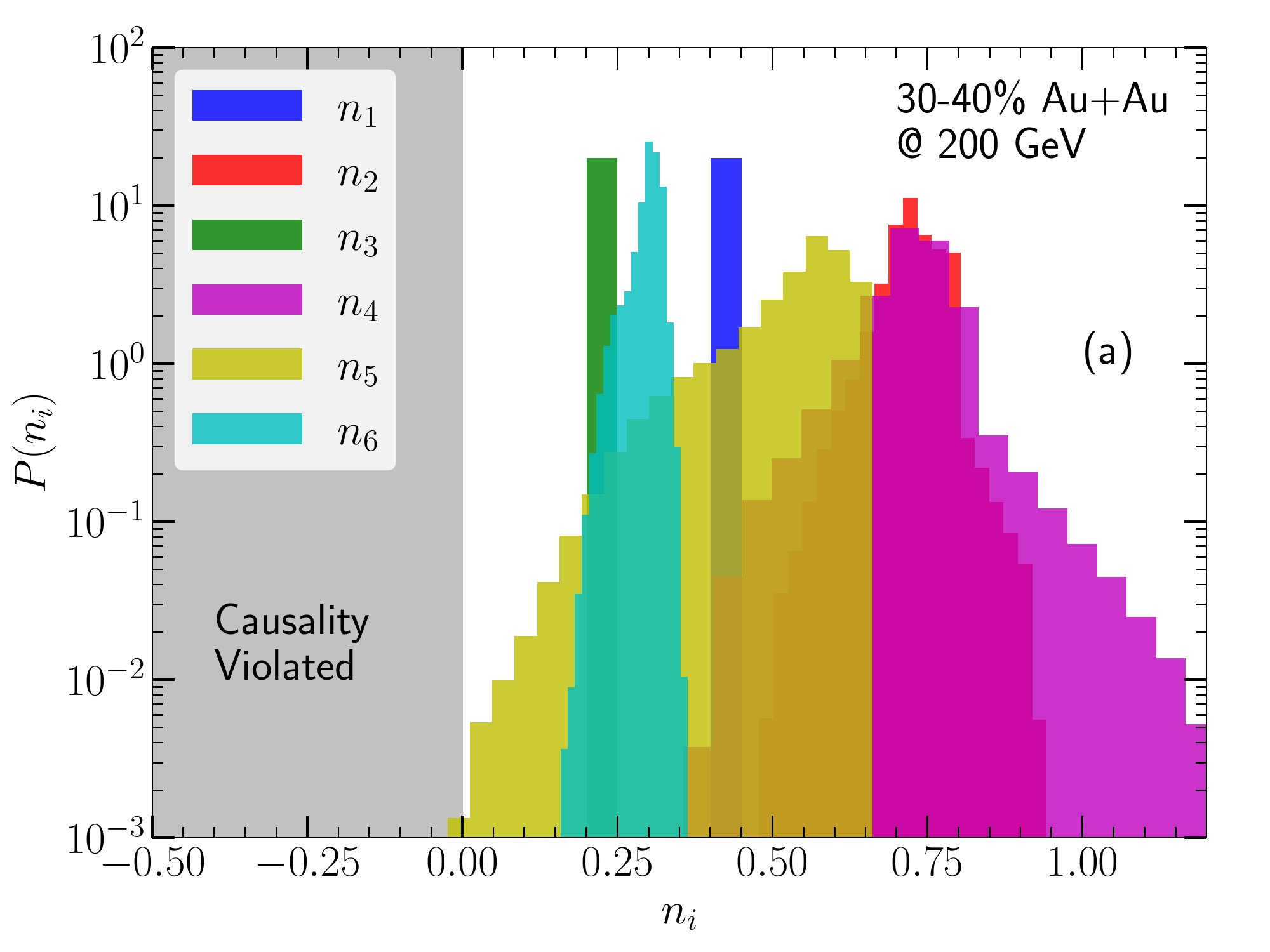}
    \includegraphics[width=0.9\linewidth]{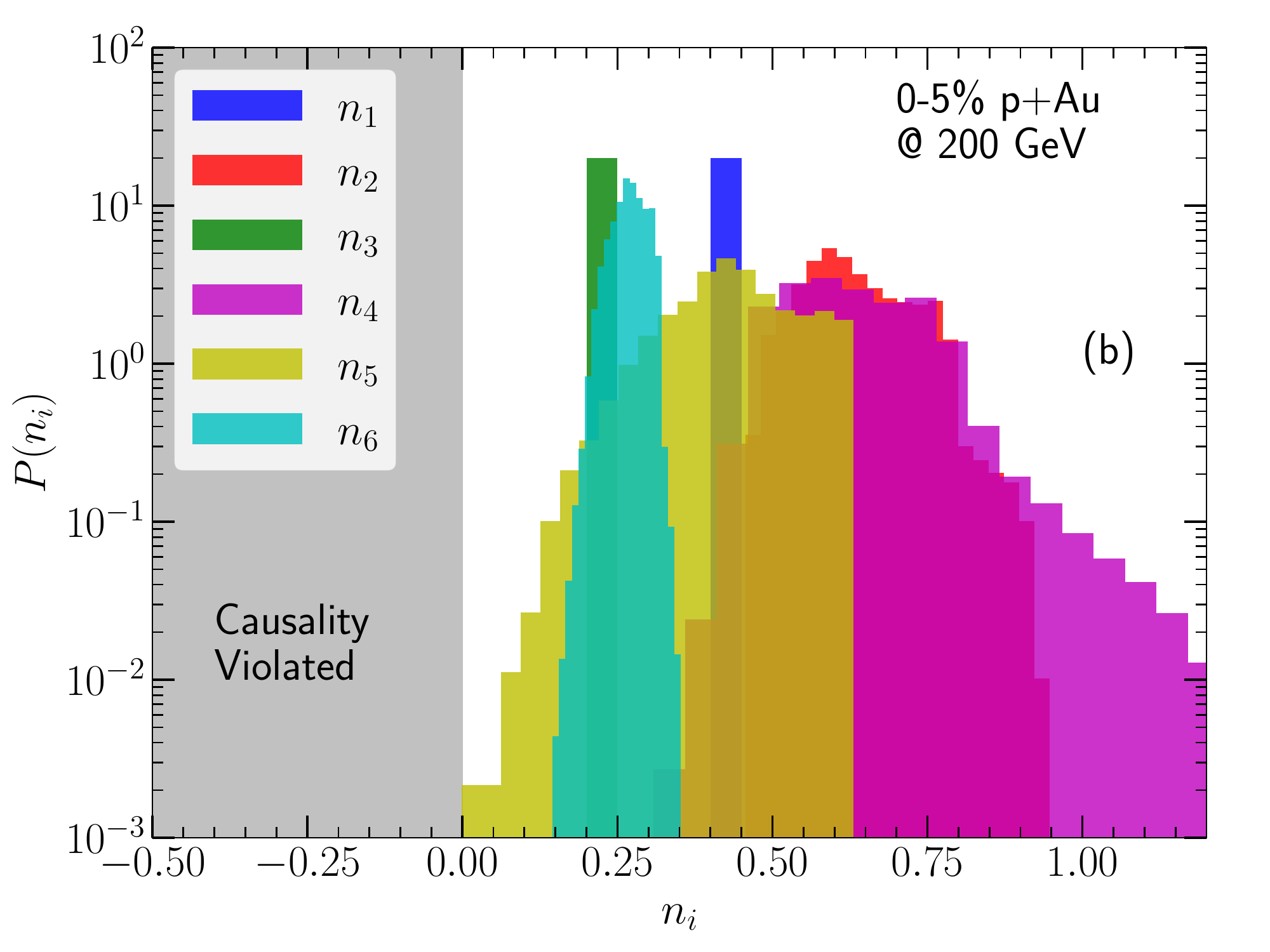}
  \caption{(Color online) Similar probability distributions as those in Fig.~\ref{fig:Pn_DNMR_tauPi1} but for hydrodynamic evolution simulated with the restricted DNMR equation of motion and the bulk relaxation time $\tau_{\Pi, 2}$. During the evolution, we restrict $R_\pi \le 0.6$ and $\vert R_\Pi \vert \le 0.6$.}
  \label{fig:Pn_IS_tauPi2}
\end{figure}

With the second choice of bulk relaxation time $\tau_{\Pi, 2}$, the necessary causality conditions allow for larger values of the inverse Reynolds numbers during the hydrodynamic simulations than those with the $\tau_{\Pi, 1}$. We find that requiring $R_\pi$ and $\vert R_\Pi \vert$ to be smaller than 0.6 can ensure all the fluid cells within the causal region, shown in Fig.~\ref{fig:Pn_IS_tauPi2}. Condition $n_5$ in Eq.~(\ref{eq:causal_n5}) imposes the dominant constraints on the inverse Reynolds numbers in this case.

We find that with the bulk relaxation time $\tau_{\Pi, 2}$, most of the fluid cells that violate the necessary causality conditions are at the first few time steps of the evolution, shown in Fig.~\ref{fig:InvReynoldsNumDis}. Because the early-stage heavy-ion collisions are far out-of-equilibrium, the necessary causality conditions set restrictions on when we can apply the relativistic viscous hydrodynamic description. Before applying the hydrodynamic framework to the system, we need to rely on effective kinetic theory to drive the system to be close enough to the local thermal equilibrium \cite{Kurkela:2018wud, Kurkela:2018vqr, Gale:2020xlg, NunesdaSilva:2020bfs}.

We note that choosing a slow varying function for coefficient $C_\zeta(c_s^2)$ allows larger viscous corrections during the hydrodynamic evolution. 
In the simulations with the bulk relaxation time $\tau_{\Pi, 1}$, a group of fluid cells near the transition region, where the square of sound speed is near 0.15, also violates the necessary causal conditions. Figure~\ref{fig:staticCausalRegion} shows that there is little room left for the dynamically evolving viscous tensor when $c_s^2 \sim 0.15$. 

\subsection{Effects of regulating viscous stress tensor with causality constraints on flow observables}

Finally, we study the effects of imposing these restrictions on the inverse Reynolds numbers on flow observables.

\begin{figure}[t!]
  \centering
    \includegraphics[width=1.0\linewidth]{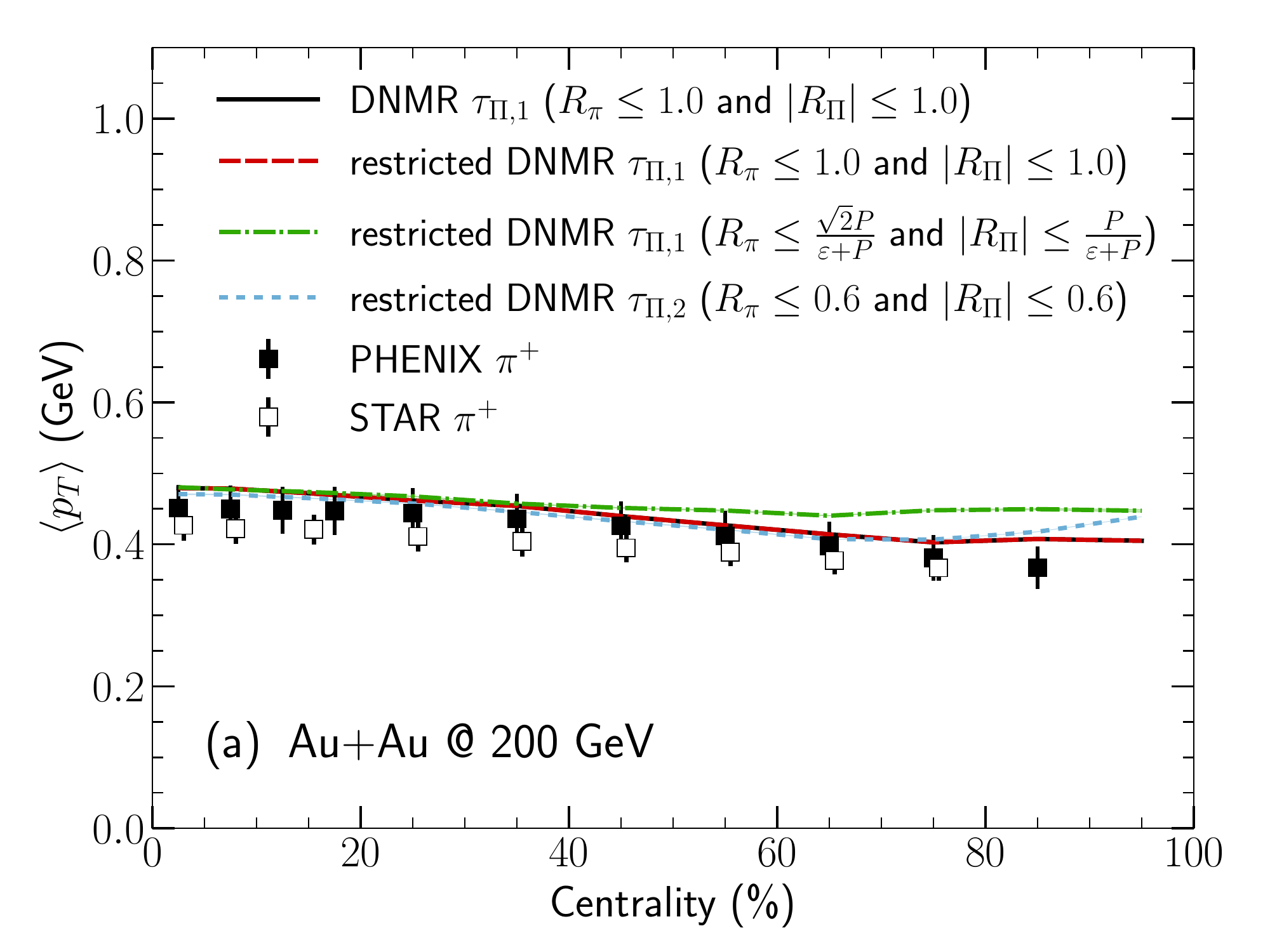} \includegraphics[width=1.0\linewidth]{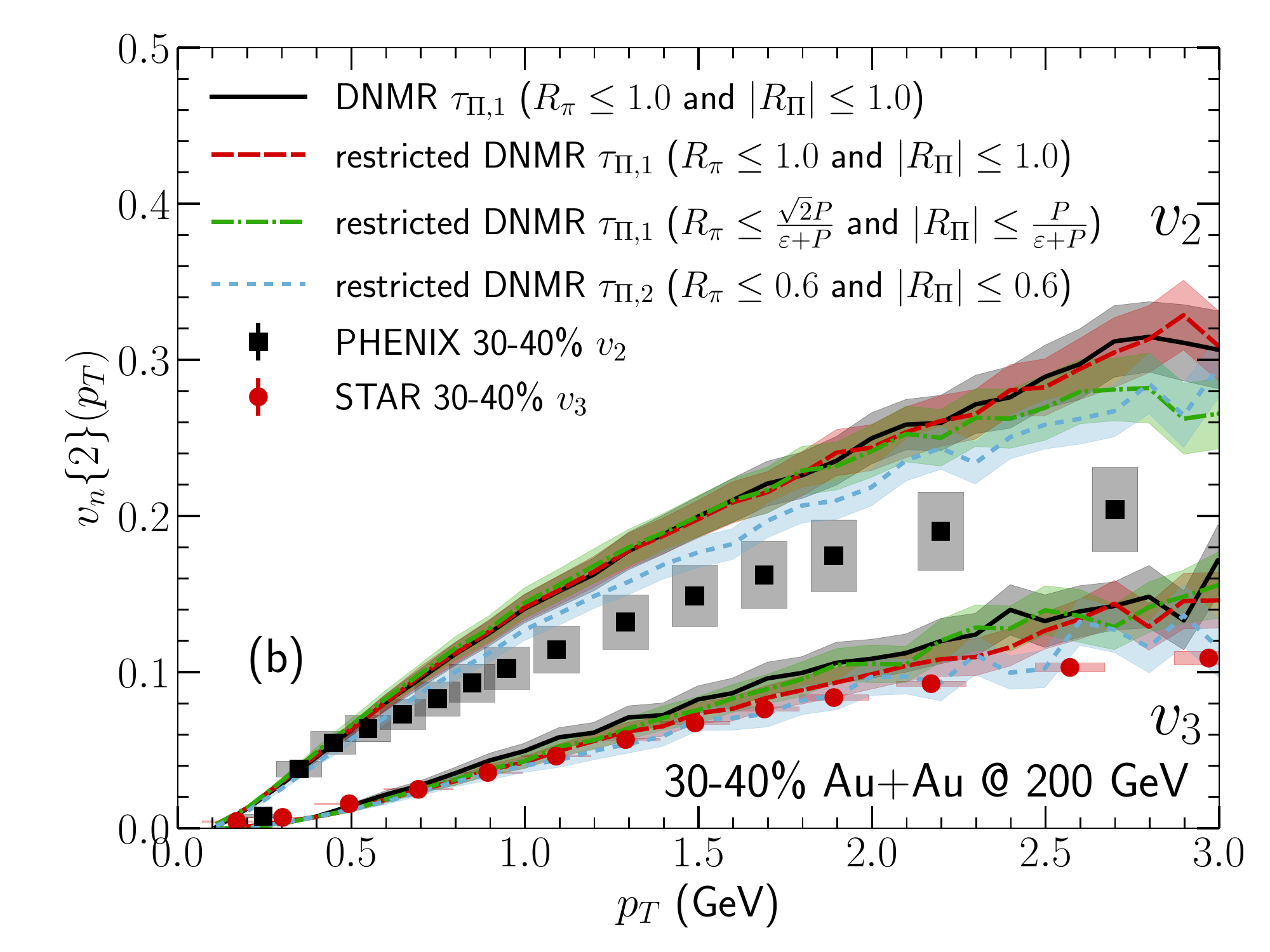}
  \caption{(Color online) The averaged transverse momentum $\langle p_T \rangle$ (a) and $p_T$-differential flow coefficients $v_n(p_T)$ (b) in Au+Au collisions at 200 GeV \cite{Adler:2003cb, Abelev:2008ab, Adare:2011tg, Adamczyk:2013waa}.}
  \label{fig:AuAuObservables}
\end{figure}

Figure~\ref{fig:AuAuObservables} shows the averaged transverse momentum $\langle p_T \rangle$ of pions compared with the PHENIX and STAR measurements in Au+Au collisions at the top RHIC energy \cite{Adler:2003cb, Abelev:2008ab} and the $p_T$-differential anisotropic flow coefficients $v_{2, 3}\{2\}(p_T)$ in 30-40\% centrality \cite{Adare:2011tg, Adamczyk:2013waa}. We first discuss the effects from the second-order transport coefficients $\{\tau_{\pi\pi}, \lambda_{\pi\Pi}, \lambda_{\Pi\pi}, \varphi_7\}$ on the flow observables. Comparing the black curves with red dashed lines, we find that these second-order terms in the hydrodynamic equations of motion have negligible effects on particle's averaged transverse momentum and $p_T$-differential anisotropic flow coefficients. Further imposing restrictions on the inverse Reynolds numbers, $R_\pi \le \sqrt{2}P/(\varepsilon + P)$ and $\vert R_\Pi \vert \le P/(\varepsilon + P)$, we ensure all the fluid cells satisfy the necessary causality conditions. We find negligible effects on the $p_T$-differential anisotropic flow coefficients. The pions' mean $p_T$ increases by 5-10\% in the peripheral centrality bins. With the second choice of the bulk relaxation time $\tau_{\Pi, 2}$, the inverse Reynolds numbers are allowed to reach up to 0.6. Comparing the flow results from the two relaxation times, we see that the elliptic flow coefficient $v_2\{2\}(p_T)$ is $\sim 10\%$ smaller with $\tau_{\Pi, 2}$ than the results from simulations with $\tau_{\Pi, 1}$ for $p_T \ge 1$\,GeV.
We checked that the additional second-order terms in the DNMR theory have less than 5\% effects on observables in the simulations with bulk relaxation time $\tau_{\Pi, 2}$.

\begin{figure}[t!]
  \centering
  \includegraphics[width=1.0\linewidth]{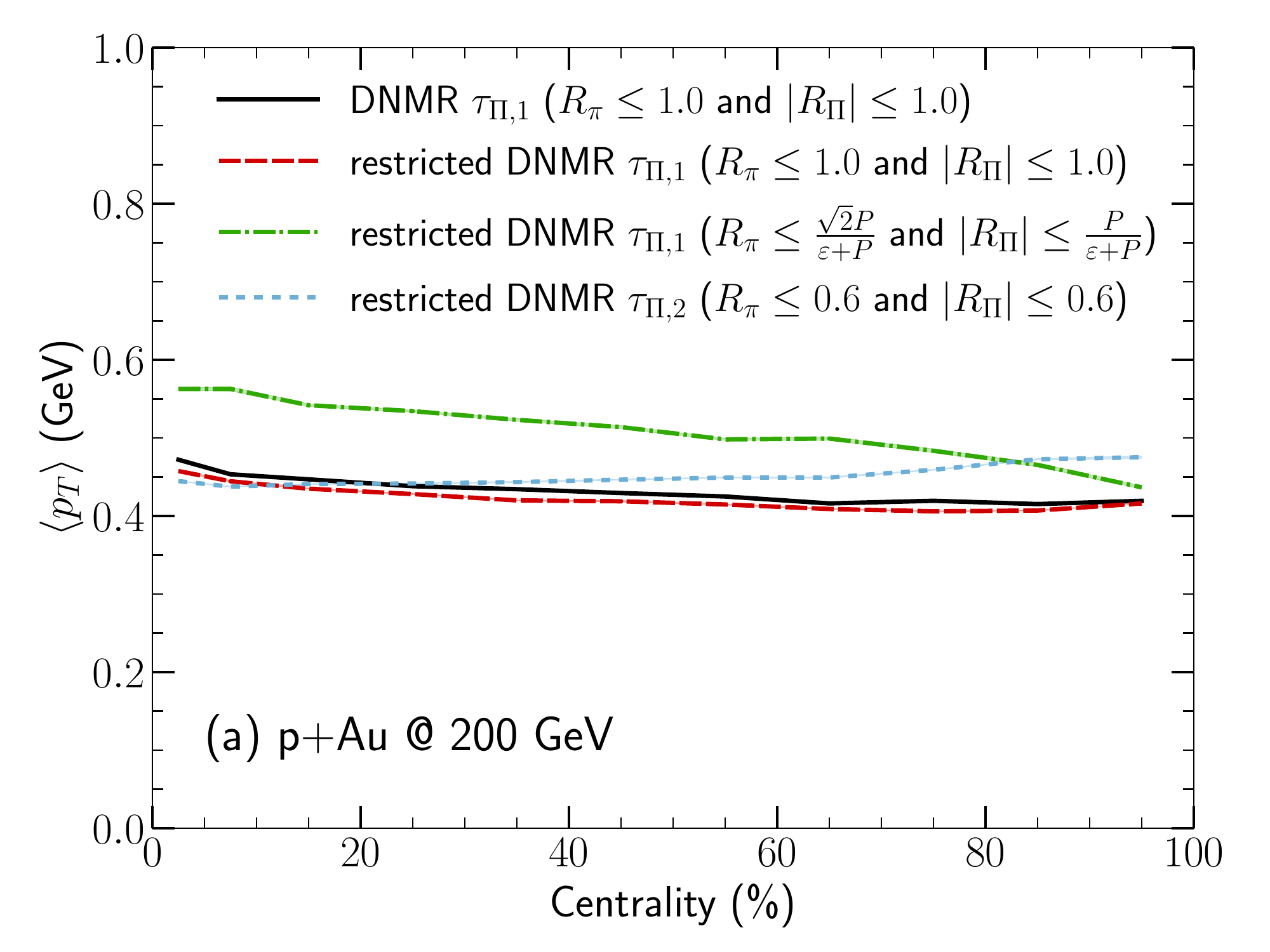} \includegraphics[width=1.0\linewidth]{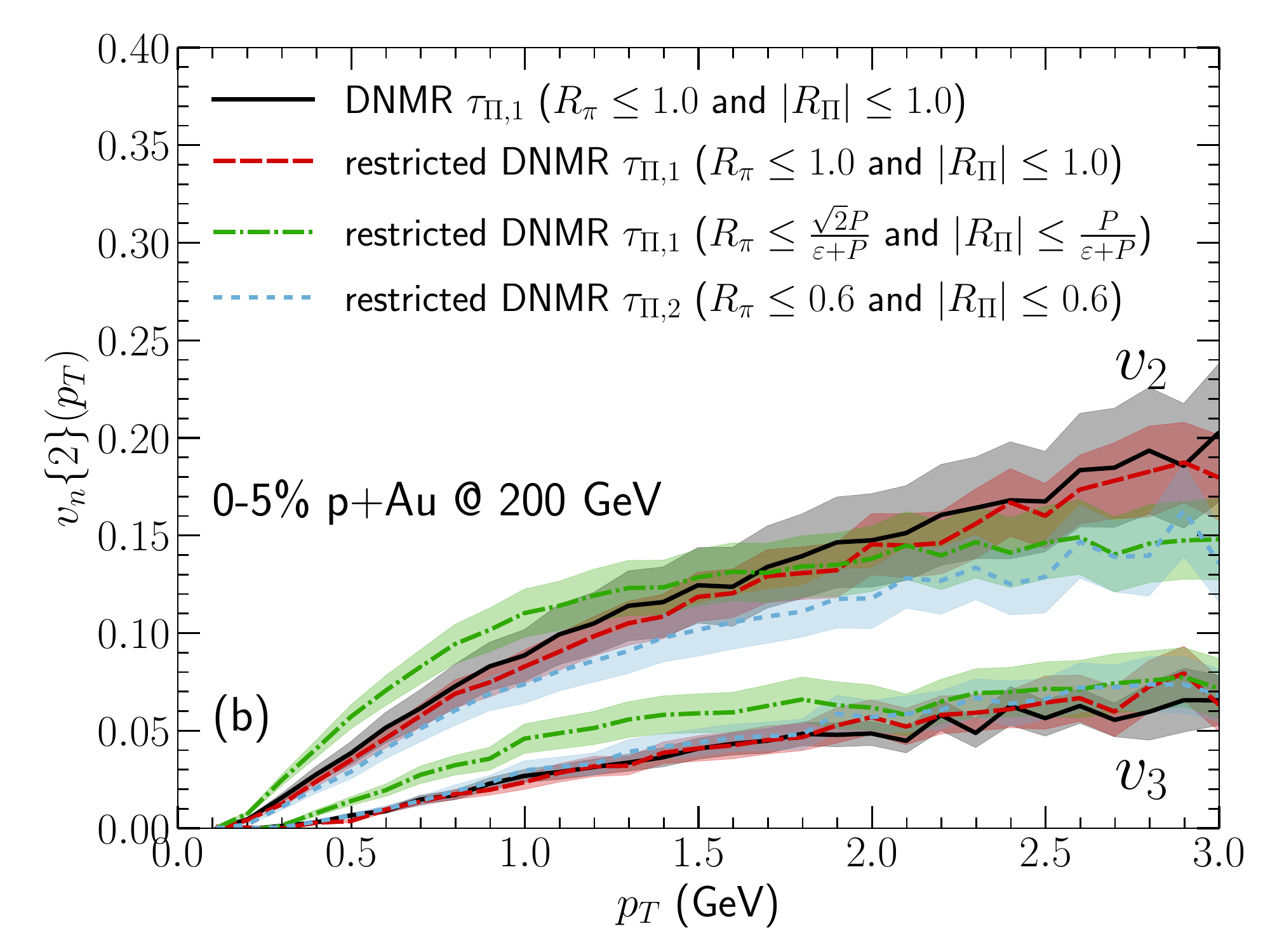}
  \caption{(Color online) The averaged transverse momentum $\langle p_T \rangle$ (a) and $p_T$-differential flow coefficients $v_n(p_T)$ (b) in p+Au collisions at 200 GeV.}
  \label{fig:pAuObservables}
\end{figure}

Figure~\ref{fig:pAuObservables} shows the same flow observables for p+Au collisions at 200 GeV. Comparing the black and red curves, we find that the effects from the additional second-order transport coefficients in the DNMR theory remain negligible in the calculations. For p+Au collisions, restricting the inverse Reynolds numbers (green dashed-dotted curves) leads to a $\sim20\%$ larger mean $p_T$ for pions. We also find that these conditions result in $20-30\%$ larger anisotropic flow coefficients $v_{2,3}$. These variations in observables are the direct consequence of restricting the viscous corrections in the simulations.
The smaller shear and bulk viscous tensors allow for stronger anisotropic and radial flow during the hydrodynamic evolution, respectively. With the second choice of the bulk relaxation time, larger viscous corrections are allowed in the hydrodynamic evolution than those simulations with $\tau_{\Pi, 1}$. With the necessary causality conditions fulfilled, the simulations with the relaxation time $\tau_{\Pi, 2}$ have $20-30\%$ smaller $v_{2,3}(p_T)$ for $p_T < 2$\,GeV than those with $\tau_{\Pi, 1}$.

\section{Conclusions} \label{sec:conc}

This work analyzes the full non-linear necessary and sufficient causality conditions \cite{Bemfica:2020xym} in the relativistic hydrodynamic description of heavy-ion collisions. We visualize the causal regions as functions of the system's inverse Reynolds number for the restricted and full DNMR hydrodynamic theories. We find the second-order transport coefficients $\{\tau_{\pi\pi}, \lambda_{\pi\Pi}, \lambda_{\Pi\pi}\}$ derived from kinetic theory with the 14-moment approximation set strong constraints on the maximum allowed inverse Reynolds numbers to satisfy causality. We explore simulations with two classes of bulk relaxation time derived from kinetic and strongly-coupled theories.

We examine the causality conditions in the hydrodynamic evolution for a typical 30-40\% Au+Au collision and a 0-5\% p+Au collision at the top RHIC energy. We find that the conditions $n_5$ and $n_6$ in Eqs.~(\ref{eq:causal_n5}) and (\ref{eq:causal_n6}) impose dominant constraints on the inverse Reynolds numbers' size. For restricted DNMR hydrodynamics with the bulk relaxation time $\tau_{\Pi, 1}$ in Eq.~(\ref{eq:tau_Pi1}), we find that $R_{\pi} \le \sqrt{2} P/(\varepsilon + P)$ and $\vert R_{\Pi} \vert \le P/(\varepsilon + P)$ can effectively ensure all fluid cells stay within the causal region. For simulations with the bulk relaxation time $\tau_{\Pi, 2}$ in Eq.~(\ref{eq:tau_Pi2}), the necessary causality conditions allow for inverse Reynolds numbers up to 0.6. Hence, larger shear and bulk viscosity can be used in hydrodynamic simulations with $\tau_{\Pi, 2}$ than those in the simulations with $\tau_{\Pi, 1}$.

We study how experimental flow observables are affected when imposing the necessary causality constraints on the inverse Reynolds numbers during hydrodynamic simulations. We find that the variations are within 10\% for the pion's mean $p_T$ and $p_T$-differential anisotropic flow coefficients for Au+Au collisions at 200 GeV. Therefore, the previous results with $R_\pi \le 1$ and $\vert R_\Pi \vert \le 1$ remain reliable, although about $4\%$ of the fluid cells violates causality. For the smaller p+Au collisions, sizable effects are present when we use the bulk relaxation time $\tau_{\Pi, 1}$ in the simulations. Restricting the size of viscous stress tensor leads to 20\% larger mean $p_T$ and $v_{2,3}(p_T)$. The bulk relaxation time $\tau_{\Pi, 2}$ allows for larger viscous corrections in numerical simulations, and the results are close to those in Ref.~\cite{Schenke:2020mbo} while ensuring all fluid cells satisfy the necessary causality conditions.

Finally, we emphasize that numerical hydrodynamic codes should build in checks for causality conditions during the evolution. They are crucial for future large-scale Bayesian Inference studies to extract the QGP transport coefficients from experimental measurements. On the one hand, the causality conditions limit the maximum allowed shear and bulk viscosity and their relaxation times in the prior of Bayesian calibration. On the other hand, our results suggest that regulating simulations with the necessary causality conditions could introduce a $20\%$ theoretical uncertainty in Bayesian analysis with flow observables in peripheral AA and pA systems.  Moreover, the causality conditions also limit when relativistic hydrodynamics can be applied at early time and request realistic pre-equilibrium dynamics. The free-streaming model used in previous Bayesian analysis \cite{Bernhard:2016tnd, Moreland:2018gsh, Bernhard:2019bmu, Everett:2020yty, Everett:2020xug, Nijs:2020roc, Nijs:2020ors} drives the collision system further away from local thermal equilibrium and increases shear stress tensor's size \cite{Liu:2015nwa}. A long free-streaming time would drive fluid cells' inverse Reynolds number $\sqrt{\pi^{\mu\nu}\pi_{\mu\nu}}/P \rightarrow \sqrt{6}/2$ \cite{Liu:2015nwa}, which increases the violation of sufficient causality conditions and the theoretical uncertainty from regulating them at the beginning of the hydrodynamic phase. Therefore, it is essential to employ a realistic pre-equilibrium evolution based on effective kinetic theories, such as K{\o}MP{\o}ST \cite{Kurkela:2018wud, Kurkela:2018vqr}, to drive the collision system to the causal region before starting hydrodynamic simulations \cite{Gale:2020xlg, NunesdaSilva:2020bfs, Plumberg:2021bme}.

\section*{Acknowledgments}
We thank Charles Gale, Ulrich Heinz, Jorge Noronha, Jean-Francois Paquet, Bjoern Schenke, and Mayank Singh for fruitful discussion. This work is supported in part by the U.S. Department of Energy (DOE) under grant number DE-SC0013460 and in part by the National Science Foundation (NSF) under grant number PHY-2012922. This research used resources of the National Energy Research Scientific Computing Center, which is supported by the Office of Science of the U.S. Department of Energy under Contract No. DE-AC02-05CH11231, resources provided by the Open Science Grid, which is supported by the National Science Foundation and the U.S. Department of Energy's Office of Science, and resources of the high performance computing services at Wayne State University. This work is supported in part by the U.S. Department of Energy, Office of Science, Office of Nuclear Physics, within the framework of the Beam Energy Scan Theory (BEST) Topical Collaboration.

\appendix*
\section{Sufficient conditions for causality}
\label{Sec:appendix}

\begin{figure}[t!]
  \centering
    \includegraphics[width=1.0\linewidth]{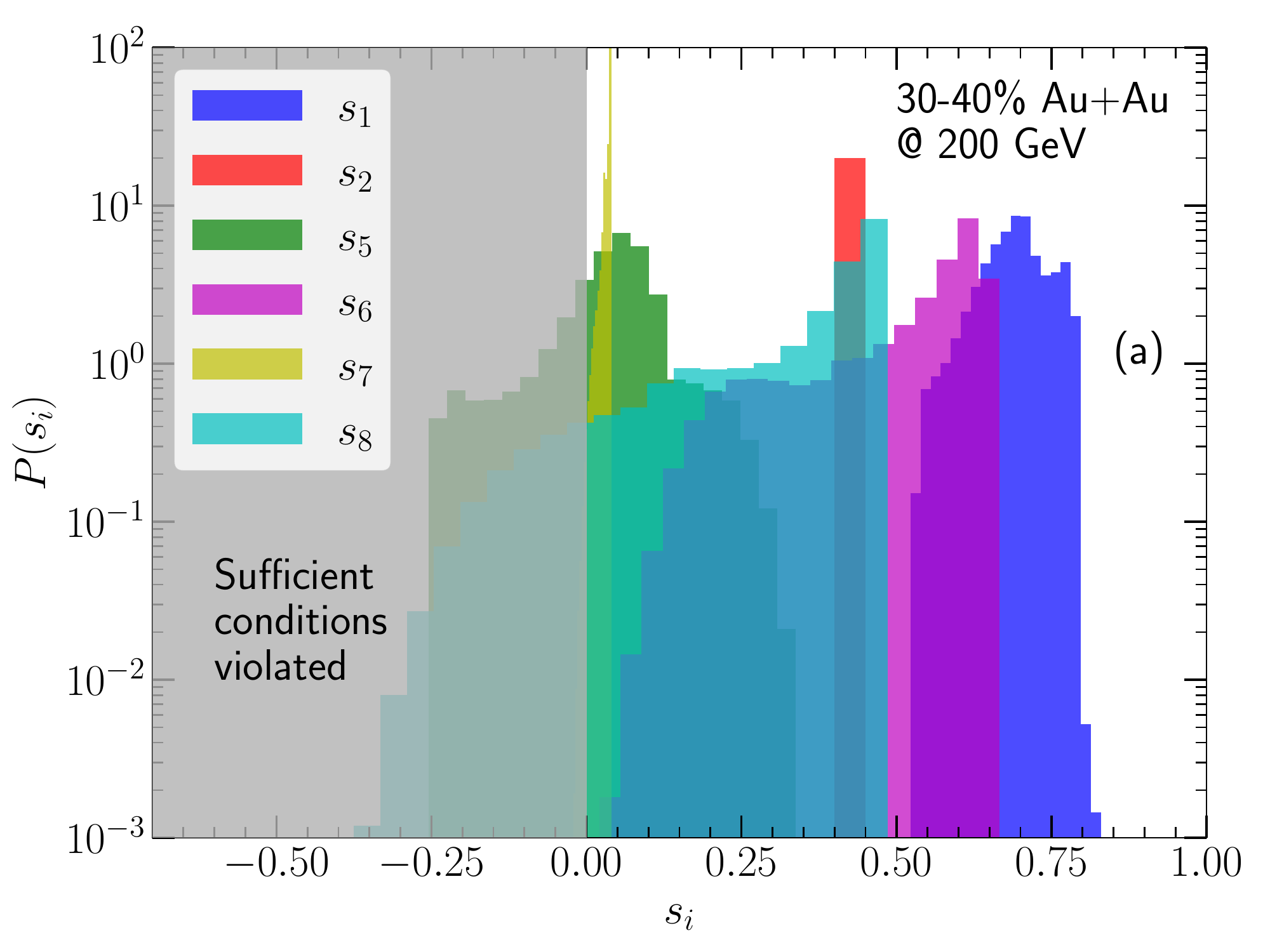}
    \includegraphics[width=1.0\linewidth]{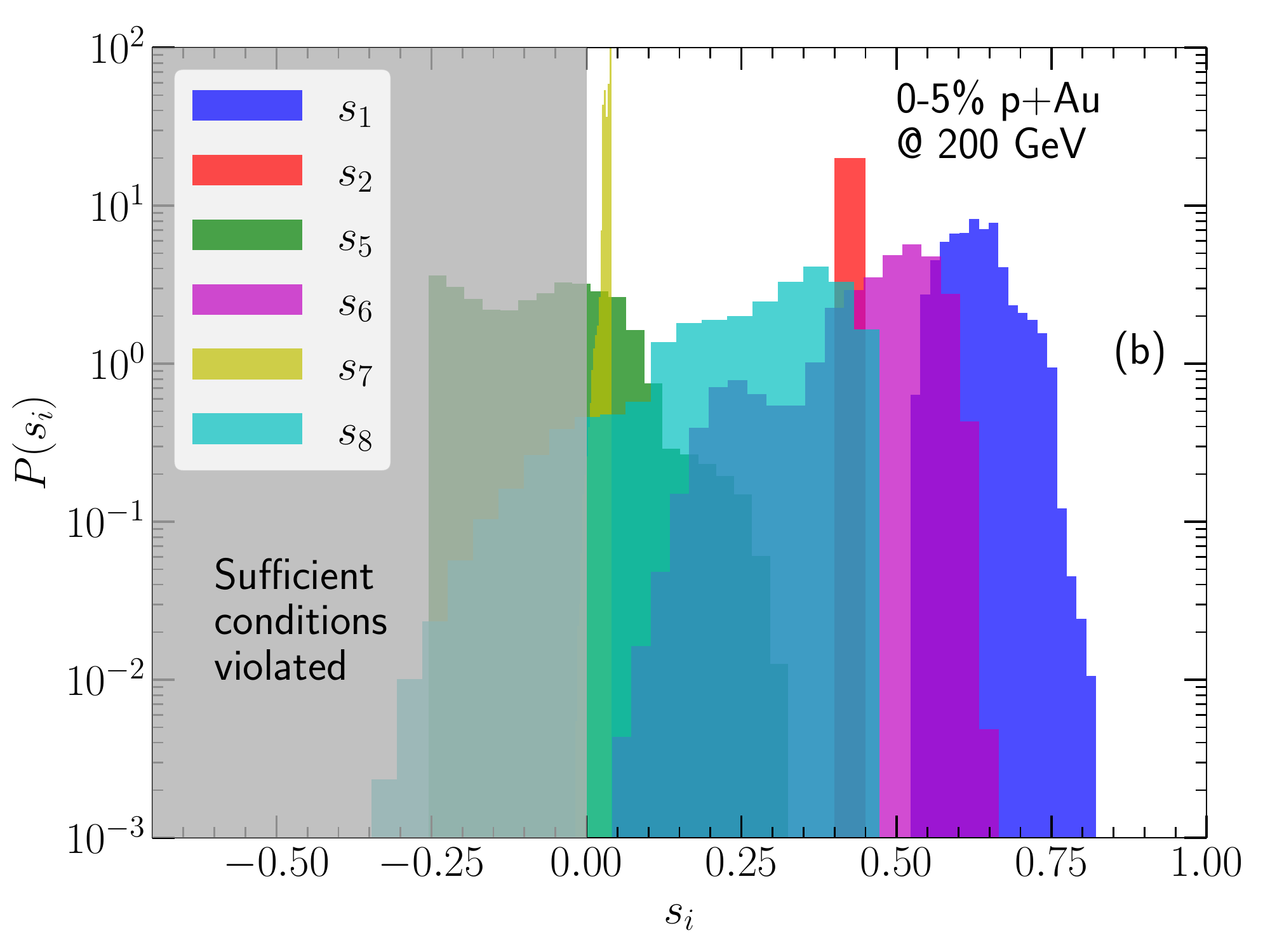}
  \caption{(Color online) Probability distributions for the sufficient causality measures in fluid cells with temperature above 145 MeV in a typical Au+Au collision at 30-40\% centrality (a) and a 0-5\% p+Au collision (b) at 200 GeV. Hydrodynamic evolution is simulated with the restricted DNMR equation of motion and the bulk relaxation time $\tau_{\Pi, 1}$. During the evolution, we restrict $R_\pi \le \sqrt{2} P/(\varepsilon + P)$ and $\vert R_\Pi \vert \le P/(\varepsilon + P)$.}
  \label{fig:Ps_IS_tauPi1}
\end{figure}

Following the Ref.~\cite{Bemfica:2020xym}, the sufficient conditions for causality can be rewritten as follows,
\begin{eqnarray}
   s_1 &\equiv& 1 - \frac{1}{C_\eta} - \frac{\vert \Lambda_1 \vert}{\varepsilon + P} + \left(1 - \frac{\lambda_{\pi\Pi}}{2\tau_\pi} \right) \frac{\Pi}{\varepsilon + P} \nonumber \\ 
   && - \frac{\tau_{\pi\pi}}{2\tau_\pi} \frac{\Lambda_3}{\varepsilon + P} \geq 0,
   \label{eq:causal_s1}
\end{eqnarray}
\begin{equation}
    s_2 \equiv \frac{1}{C_\eta} + \frac{\lambda_{\pi\Pi}}{2\tau_\pi} \frac{\Pi}{\varepsilon + P} - \frac{\tau_{\pi\pi}}{2\tau_\pi} \frac{\vert \Lambda_1 \vert}{\varepsilon + P} \geq 0,
    \label{eq:causal_s2}
\end{equation}
\begin{equation}
    s_3 \equiv 6 \frac{\delta_{\pi\pi}}{\tau_\pi} - \frac{\tau_{\pi\pi}}{\tau_\pi} \geq 0,
    \label{eq:causal_s3}
\end{equation}
\begin{equation}
    s_4 \equiv \frac{\lambda_{\Pi\pi}}{\tau_\Pi} + c^2_s-\frac{\tau_{\pi\pi}}{12\tau_\pi} \geq 0,
    \label{eq:causal_s4}
\end{equation}
\begin{eqnarray}
    s_5 &\equiv& \left(1 + \frac{\Pi}{\varepsilon + P}\right)(1 - c_s^2) \nonumber \\
    && - \bigg[ \frac{4}{3} \frac{1}{C_\eta} + \frac{1}{C_\zeta} + \left(\frac{2}{3} \frac{\lambda_{\pi\Pi}} {\tau_\pi} +  \frac{\delta_{\Pi\Pi}}{\tau_\Pi}\right) \frac{\Pi}{\varepsilon + P} \nonumber \\ 
    && + \left(\frac{\delta_{\pi\pi}}{\tau_\pi} + \frac{\tau_{\pi\pi}}{3 \tau_\pi} + \frac{\lambda_{\Pi\pi}}{\tau_\Pi} + c_s^2 \right) \frac{\Lambda_3}{\varepsilon + P} + \frac{\vert \Lambda_1 \vert}{\varepsilon + P} \nonumber \\
    && +\frac{ (\frac{\delta_{\pi\pi}}{\tau_\pi} - \frac{\tau_{\pi\pi}}{12\tau_\pi}) (\frac{\lambda_{\Pi\pi}}{\tau_\Pi} + c_s^2 - \frac{\tau_{\pi\pi}}{12\tau_\pi}) (\frac{\Lambda_3}{\varepsilon + P} + \frac{\vert \Lambda_1 \vert}{\varepsilon + P})^2}{1 - \frac{1}{C_\eta} + (1 - \frac{\lambda_{\pi\Pi}}{2 \tau_\pi}) \frac{\Pi}{\varepsilon + P} - \frac{\vert \Lambda_1 \vert}{\varepsilon + P} - \frac{\tau_{\pi\pi}}{2\tau_\pi} \frac{\Lambda_3}{\varepsilon + P} } \bigg] \nonumber \\
    && \geq 0,
    \label{eq:causal_s5}
\end{eqnarray}
\begin{eqnarray}
    s_6 &\equiv& \frac{1}{3C_\eta}  + \frac{1}{C_\zeta} + c^2_s + \left(\frac{\lambda_{\pi\Pi}}{6 \tau_\pi} + \frac{\delta_{\Pi\Pi}}{\tau_\Pi} + c_s^2 \right) \frac{\Pi}{\varepsilon + P} \nonumber \\
    && + \left(\frac{\tau_{\pi\pi}}{6 \tau_\pi} - \frac{\delta_{\pi\pi}}{\tau_\pi} + \frac{\lambda_{\Pi \pi}}{\tau_\Pi} - c_s^2 \right) \frac{\vert \Lambda_1 \vert}{\varepsilon + P} \geq 0,
    \label{eq:causal_s6}
\end{eqnarray}
\begin{eqnarray}
    s_7 &\equiv& \left[\frac{1}{C_\eta} + \frac{\lambda_{\pi\Pi}}{2\tau_\pi} \frac{\Pi}{\varepsilon + P}  - \frac{\tau_{\pi\pi}}{2\tau_\pi} \frac{\vert \Lambda_1 \vert}{\varepsilon + P} \right]^2 \nonumber \\
    && - \left(\frac{\delta_{\pi\pi}}{\tau_\pi} - \frac{\tau_{\pi\pi}}{12\tau_\pi} \right) \left(\frac{\lambda_{\Pi\pi}}{\tau_\Pi} + c_s^2 - \frac{\tau_{\pi\pi}}{12\tau_\pi} \right) \nonumber \\
    && \qquad \times \left(\frac{\Lambda_3}{\varepsilon + P} + \frac{\vert \Lambda_1 \vert}{\varepsilon + P} \right)^2 \geq 0,
    \label{eq:causal_s7}
\end{eqnarray}
\begin{eqnarray}
    s_8 &\equiv& \frac{4}{3 C_\eta} + \frac{1}{C_\zeta} + c_s^2 + \left(\frac{2}{3} \frac{\lambda_{\pi\Pi}}{\tau_\pi} + \frac{\delta_{\Pi\Pi}}{\tau_\Pi} + c_s^2 \right) \frac{\Pi}{\varepsilon + P} \nonumber \\
    && - \left(\frac{\delta_{\pi\pi}}{\tau_\pi} + \frac{\tau_{\pi\pi}}{3\tau_\pi} - \frac{\lambda_{\Pi\pi}}{\tau_\Pi} + c_s^2 \right) \frac{\vert \Lambda_1 \vert}{\varepsilon + P}  \nonumber \\
    && - \frac{(1 + \frac{\Pi}{\varepsilon + P} + \frac{\Lambda_2}{\varepsilon + P})(1 + \frac{\Pi}{\varepsilon + P} + \frac{\Lambda_3}{\varepsilon + P})}{3 \left(1 + \frac{\Pi}{\varepsilon + P} - \frac{\vert \Lambda_1 \vert}{\varepsilon + P} \right)^2} \nonumber \\
    && \quad \times \bigg[1 + \frac{2}{C_\eta} + \left(1 + \frac{\lambda_{\pi\Pi}}{\tau_\pi} \right) \frac{\Pi}{\varepsilon + P} \nonumber \\
    && \qquad - \frac{\vert \Lambda_1 \vert}{\varepsilon + P} + \frac{\tau_{\pi\pi}}{\tau_\pi} \frac{\Lambda_3}{\varepsilon + P} \bigg] \geq 0.
    \label{eq:causal_s8}
\end{eqnarray}
Because the conditions $s_3$ and $s_4$ do not depend on the dynamical evolution shear stress tensor and bulk viscous pressure, we do not need to check them during the hydrodynamic evolution.

In Fig.~\ref{fig:Ps_IS_tauPi1}, we show the probability distributions of sufficient causality conditions' measures for 30-40\% Au+Au and 0-5\% p+Au collisions at 200 GeV. Most of the violating fluid cells fail the conditions $s_5$ and $s_8$ in Eqs.~(\ref{eq:causal_s5}) and (\ref{eq:causal_s8}). 
Here, we already restrict the inverse Reynolds number $R_\pi \le \sqrt{2} P/(\varepsilon + P)$ and $\vert R_\Pi \vert \le P/(\varepsilon + P)$ to ensure all the fluid cells fulfill the necessary causality conditions.  
However, these restrictions on $R_\pi$ and $R_\Pi$ do not reduce the fractions of fluid cells that violate the sufficient conditions.

\bibliography{Causality_Conditions}

\end{document}